\documentclass[preprintnumbers, floatfix, preprintnumbers, letterpaper, superscriptaddress,nofootinbib]{revtex4}
\usepackage{graphicx}
\usepackage{microtype}
\usepackage{amsmath}
\usepackage{amssymb}
\usepackage{subfigure}
\usepackage{hyperref}
\usepackage{url}
\usepackage{xcolor}
\usepackage{color}
\usepackage{mathrsfs}
\usepackage{calrsfs}
\usepackage{amsfonts}
\usepackage{eufrak}
\usepackage{tabularx}
\usepackage{eucal}
\usepackage{latexsym}
\usepackage{ragged2e}
\usepackage{epsfig}
\usepackage{textcomp}

\usepackage{caption}
\DeclareCaptionJustification{justified}{\leftskip=0pt \rightskip=0pt \parfillskip=0pt plus 1fil}
\definecolor{vividviolet}{rgb}{0.62, 0.0, 1.0}
\definecolor{amaranth}{rgb}{0.9, 0.17, 0.31}
\definecolor{palatinateblue}{rgb}{0.15, 0.23, 0.89}
\definecolor{brightpink}{rgb}{1.0, 0.0, 0.5}
\definecolor{cornflowerblue}{rgb}{0.39, 0.58, 0.93}
\definecolor{deepcarminepink}{rgb}{0.94, 0.19, 0.22}
\definecolor{radicalred}{rgb}{1.0, 0.21, 0.37}

\hypersetup{ linktoc=all,
    colorlinks, linkcolor={palatinateblue},
    citecolor={brightpink}, urlcolor={amaranth}
}

\graphicspath{{Images/}}

\renewcommand{\d}[1]{\ensuremath{\operatorname{d}\!{#1}}}

\graphicspath{{Images/}}
\renewcommand{\d}[1]{\ensuremath{\operatorname{d}\!{#1}}}
\makeatletter
\def\@fnsymbol#1{\ensuremath{\ifcase#1\or \ddagger \or  $\textleaf$  \or \dagger
\else\@ctrerr\fi}}%

\makeatother
\def\sideremark#1{\ifvmode\leavevmode\fi\vadjust{\vbox to0pt{\vss% the remark
 \hbox to 0pt{\hskip\hsize\hskip1em%                          will appear only
 \vbox{\hsize1.3cm\tiny\raggedright\pretolerance10000%          on the side
 \noindent #1\hfill}\hss}\vbox to8pt{\vfil}\vss}}}%
\def\beq{\begin{equation}}
\def\eeq{\end{equation}}

\begin{document}

\title{The Horizon of the McVittie Black Hole: On the Role of the Cosmic Fluid Modeling}

\author{Daniele \surname{Gregoris}}
\email{danielegregoris@libero.it}
\affiliation{Center for Gravitation and Cosmology, College of Physical Science and Technology, Yangzhou University, \\180 Siwangting Road, Yangzhou City, Jiangsu Province, P.R. China 225002}
\affiliation{School of Aeronautics and Astronautics, Shanghai Jiao Tong University, Shanghai 200240, China}

\author{Yen Chin \surname{Ong}}
\email{ycong@yzu.edu.cn}
\affiliation{Center for Gravitation and Cosmology, College of Physical Science and Technology, Yangzhou University, \\180 Siwangting Road, Yangzhou City, Jiangsu Province, P.R. China 225002}
\affiliation{School of Aeronautics and Astronautics, Shanghai Jiao Tong University, Shanghai 200240, China}

\author{Bin \surname{Wang}}
\email{wang_b@sjtu.edu.cn}
\affiliation{Center for Gravitation and Cosmology, College of Physical Science and Technology, Yangzhou University, \\180 Siwangting Road, Yangzhou City, Jiangsu Province, P.R. China 225002}
\affiliation{School of Aeronautics and Astronautics, Shanghai Jiao Tong University, Shanghai 200240, China}

\begin{abstract}
In this paper,  we investigate the existence and time evolution of the cosmological and event horizons in a McVittie universe whose expansion is driven by the Redlich-Kwong, (Modified) Berthelot, Dieterici, and Peng-Robinson fluids, respectively. The equations of state of these fluids are rich enough to account for both exotic and regular, as well as ideal and non-ideal matter contents of the universe. We show that the cosmological horizon is expanding, while the event horizon is shrinking along the cosmic time evolution. The former achieves larger size for regular types of matter, contrary to the latter. The strength of interactions within the cosmic fluid are shown to play a more important role in affecting the evolution of the event horizon, rather than of the cosmological horizon in the case of a singularity-free universe. While the cosmological horizon always exists during the time evolution, the event horizon can exist only when a certain relationship between the Hawking-Hayward quasi-local mass and the Hubble function is fulfilled. In this manner, we can study the role played by the large-scale physics (cosmic evolution) on the local scale physics (evolution of a black hole).
\end{abstract}
\maketitle

\section{Introduction}

Our Universe is populated by many different objects: stars, galaxies, clusters of galaxies, black holes, and possibly dark energy. There is no reason for postulating that all these astrophysical entities should evolve independently from each other without experiencing feedback phenomena and energy exchanges. For example, the molecules constituting the cosmic fluid or a nearby star are likely to be attracted gravitationally by a black hole with the possibility of falling into it and contributing to its accretion or to its evaporation. Therefore,  the most commonly adopted Schwarzschild or Kerr metrics for a static or rotating black hole  (even their generalizations that include a cosmological constant), respectively, seem unsatisfactory for a realistic picture of an astrophysical black hole for many reasons which essentially follow from the two mathematical assumptions on which they are based: they do not allow the black hole to evolve and change in time because of the hypothesis of stationarity, and they are vacuum solutions of the gravitational field equations of general relativity. 

Neglecting the presence of matter in its proximity, the Kerr solution cannot provide any possible mechanism for the formation of an accretion disk around the central massive source \cite{disk1,disk2}, whose existence instead has been established from the study of active galactic nuclei \cite{agn} and of the jets emitted by the particles falling into it \cite{jet}. Moreover, its time-independence does not allow us to track the stages of the formation of the black hole under the gravitational collapse of a star \cite{tov1,tov2,tov3}, to follow its growth by accretion of interstellar gas \cite{tov4,tov5,tov6}, and to account for its possible ``explosion'' at the final stage of Hawking evaporation and disappearance thereafter \cite{tov7}.

Most importantly, the Kerr solution does not take into account that the black hole should live inside a Universe which, according to the current standard model of cosmology ($\Lambda$-Cold Dark Matter model or $\Lambda$CDM in short), expand in time and it is not empty but dominated by some form of dark energy which is taken to be a cosmological constant in the simplest scenario \cite{peebles}. Along this line of thinking, one can adopt the Schwarzschild-(anti-)de Sitter metric for describing a spherically symmetric black hole embedded in a spacetime whose matter-energy is given by the cosmological constant \cite{or3}. In this latter case the spacetime is no longer asymptotically flat, and the location of the event horizon is shifted by the presence of the cosmological constant. However, the most recent estimates of the cosmological parameters from the Planck mission suggest the possibility that the equation of state for dark energy is evolving in time, ruling out the possibility of modeling it as a cosmological constant within 1$\sigma$ at the 95\% confidence level \cite{planck}. Taking into account this requirement, it is conceivable that modeling a black hole through the McVittie metric will constitute an improvement because it allows for a simultaneous time evolution of the mass of the black hole and of the scale factor of the Universe under a generic matter content \cite{1933}.  

For dealing with a well posed problem in the framework of general relativity, after fixing the geometrical symmetries of the configuration it is necessary to specify what is the type of matter entering the stress-energy tensor appearing in the Einstein's equations. In the light of the aforementioned cosmological discussion, after choosing the McVittie manifold, we propose to picture dark energy following the dynamical models due to Redlich-Kwong \cite{reos1}, (Modified) Berthelot \cite{reos2}, Dieterici \cite{reos3}, and Peng-Robinson \cite{reos4} because their astrophysical applicability as equations of state for dark fluids has been already tested in \cite{capo}. The purpose of this manuscript is to quantify how the location, time evolution and size of the McVittie horizon are affected by these four different modelings for the dark energy fluid, and to clarify the differences with respect to the previously adopted  Schwarzschild-(anti-)de Sitter metric. It will be shown that the cosmological horizon is expanding in time, while the event horizon is shrinking. The role of the matter content exotic vs. regular, and ideal vs. non-ideal\footnote{Following the language of \cite{cosmo,poly}, we name {\it exotic matter} a fluid supported by a negative pressure, {\it regular matter} a fluid supported by a positive pressure, {\it ideal matter} a fluid whose pressure and energy density are related to each other through a linear functional, and {\it non-ideal matter} a fluid whose pressure and energy density are related through a non-linear law. For example, the equation of state $p=a \rho^b$ describes an exotic ideal matter for $a<0$ and $b=1$, an exotic non-ideal matter for $a<0$ and $b \neq 1$, a regular ideal matter for $a>0$ and $b=1$, and finally a regular non-ideal matter for $a>0$ and $b\neq 1$. }  is analyzed showing that the cosmological horizon is larger for regular types of matter contrary to the event horizon. It also turns out that the strength of interactions within the fluid plays a more important role in the evolution of the event horizon rather than for the cosmological horizon in a singularity-free universe. On the other hand, a singularity in the pressure can be realized for certain values of these interactions in two of the equations of state that we studied; this divergence strongly affects the evolution of both the cosmological and event horizons.

Our manuscript is organized as follows: in sec. (\ref{sII}) we introduce the McVittyie spacetime as a solution of the Einstein's equations of general relativity, in particular relating the evolution of the mass of the central object to the expansion of the whole universe through the Hawking-Hayward quasi-local mass, and deriving the time-dependent positions of the horizons in terms of an algebraic third-degree equation; the occurrence of one or more real root according to the interplay of the Hubble function and of the Hawking-Hayward quasi-local mass is discussed. Then, in sec. (\ref{sIII}) we introduce the four models for the cosmic fluid that we shall investigate in this paper, which can account both for exotic or regular and for ideal or non-ideal matter contents, for example ranging from dark energy to massless scalar fields and pressure-less dark matter. Also, in this section we derive the time evolution for the Hubble function in a McVittie universe. Sec. (\ref{sIV}) exhibits the numerical analysis of the time evolution of the cosmological and of the event horizon of the McVittie spacetime in terms of a number of plots whose common properties and differences are then commented in light of the equations of state we adopted. Finally we draw out our conclusion in sec. (\ref{sV}), commenting on how our work fits in the research lines trying to connect large-scale and local-scale physics in our Universe.

\section{The McVittie horizon: setup of the problem} \label{sII}

In an isotropic coordinate system $x^\mu=$($t$, $r$, $\theta$, $\phi$) and in units such that $c=1={8\pi G}$, the McVittie metric reads as \cite{1933}:
\beq
\label{metric}
\d s^2= -\frac{\left(  1-\frac{m(t)}{2r}  \right)^2}{\left(  1+\frac{m(t)}{2r}  \right)^2}\d t^2 +a^2(t) \left(1+\frac{m(t)}{2r}\right)^4 (\d r^2 +r^2\d\theta^2 + r^2 \sin^2 \theta \d\phi^2)\,.
\eeq
It describes a black hole of mass $m(t)$ embedded in a Friedmann-like universe with scale factor $a(t)$ allowing for matter-energy exchanges between the central massive source and the surrounding space. 
Note that if $m(t)=\text{const.}$ and $a(t)\equiv 1$, then we recover the standard Schwarzschild metric in the isotropic coordinates.

In what follows we may simply denote $m=m(t)$, and $a=a(t)$ keeping in mind their time dependence. The McVittie metric can be interpreted as the generalization of the Schwarzschild-(anti-)de Sitter spacetime because it allows for a more general time evolution of the scale factor of the universe which depends on the type of matter driving its expansion beyond the simplest cosmological constant scenario. This time evolution is accounted for by the Einstein field equation $G_{\mu\nu}=T_{\mu\nu}$ and the Bianchi identities $T^{\mu\nu}{}_{;\nu}=0$, in which a semicolon  denotes a covariant derivative, $G_{\mu\nu}$ is the Einstein tensor,  $T_{\mu \nu}=(\rho +p) u_\mu u_\nu +p g_{\mu\nu} $ is the stress-energy tensor denoting the matter content of the universe in terms of its energy density $\rho=\rho(t,r)$, its pressure $p=p(t,r)$, and the four-velocity of the reference free-falling observer $u^\mu=\frac{1}{\sqrt{-g_{tt}}}\delta_t^\mu$.
McVittie himself imposed a closure relation between the mass of the black hole and the scale factor of the universe as:
\beq
\frac{\dot m}{m}=-\frac{\dot a}{a}\,,
\eeq
where an over dot denotes a derivative with respect to time. This assumption avoids a nontrivial $G^{t}{}_{r}$ component of the field equations, and so the previous equation can be integrated into
\beq
\label{m(t)}
m(t)=\frac{m_H}{a(t)}\,,
\eeq
where the constant of integration $m_H$ has been later interpreted in the literature as the Hawking-Hayward quasi-local mass \cite{mass1,mass2,mass3,mass4}. The Hawking-Hayward quasi-local mass arose in the debate of a possible definition for what gravitational energy is in general relativity, which must be well defined on the horizon of a black hole (because it is a compact orientable spatial 2-surface), and which must reduce to the Arnowitt-Deser-Misner (ADM) mass at spatial infinity \cite{ADM}.

So far, the literature has already explained how to find the 2-surface that gives the horizon of this black hole when the expansion of the universe is driven by a mixture of non-interacting pressure-less dark matter and a cosmological constant \cite{faraoni1,faraoni2,faraoni3}. Moreover, a procedure for locating the McVittie horizon in terms of the zeros of an appropriate curvature invariant has been proposed along the research which has been trying to develop local techniques for detecting a black hole horizon \cite{page}. Those algorithms do not rely on the non-local propagation of light rays and  constitute the ground for the {\it geometric horizon conjecture} \cite{loc1,loc2,loc3,loc4,loc5,loc6,loc7}. In this paper we are interested in understanding how the horizon actually looks like when specific and different matter contents of the universe are considered. In particular, we are interested in extending the case of the Schwarzschild-(anti-)de Sitter black hole replacing the cosmological constant with a two-parameter non-ideal equation of state for the dark energy fluid. Therefore, we postpone the analysis of the effect of the presence of dark matter, possibly interacting with dark energy, to a future study. 

The horizon of the McVittie black hole can be found by imposing the condition $||\nabla \tilde r||^2=0$, where
\beq
\tilde r= a(t) \left(1+\frac{m(t)}{2r}\right)^2 r
\eeq
is the areal radius \cite{ash}. Explicitly we get the following algebraic equation:
\beq
\frac{  \left[\left(a(t)r+\frac{m_H}{2}\right)^3 \dot a(t) +r a^2(t)\left(a(t)r-\frac{m_H}{2}\right)  \right] \left[ r a^2(t)\left(a(t)r-\frac{m_H}{2}\right)-\left(a(t)r+\frac{m_H}{2}\right)^3 \dot a(t)  \right] }{ \left(a(t)r+\frac{m_H}{2}\right)^2 r^2 a^4(t)   }=0               \,.
\eeq
This condition can be recast into
\beq
\left(\chi+\frac{m_H}{2}\right)^6 H^2(t)-\chi^2 \left(\chi-\frac{m_H}{2}\right)^2=0\,,
\eeq
in which we introduced the Friedmann comoving distance $\chi=\chi(t,r):=a(t)r$, and the Hubble function $H=H(t):=\dot a/a$ \cite{faraoni3}. Since $\chi(t)>0$ $\forall t$, we can move from a 6th order to a 3rd order algebraic equation 
\beq
\label{3rd}
\left(\chi+\frac{m_H}{2}\right)^3 H(t)=\chi \left|\chi-\frac{m_H}{2}\right|\,.
\eeq
For an expanding universe (i.e. supported by a positive Hubble function) the left hand side is positive definite, hence the absolute value sign. Thus, our procedure is based on the solution of an algebraic equation and it  is computationally more convenient than solving the differential equation which accounts for the focusing properties of a congruence of light rays as considered in \cite{davis}.  To understand the number of roots we expect for the locations of the horizons, we rewrite Eq.(\ref{3rd}) in canonical form as
\beq
\label{cgen}
A \chi^3+B \chi^2+C \chi+D=0\,,
\eeq
where the explicit expressions for the coefficients are
\beq
\label{coeff}
A=H(t)\,, \quad B= \frac{3H(t) m_H}{2}\mp 1 \,, \quad C=\frac{3H(t) m_H^2}{4} \pm \frac{m_H}{2} \,, \quad D= \frac{H(t) m_H^3}{8}\,.
\eeq
Then, the discriminant of this cubic equation is
\beq
\Delta=18ABCD-4B^3D+B^2C^2-4AC^3-27A^2D^2=\frac{m_H^2}{4}(1-27m_H^2 H^2)\,,
\eeq
in which the same result holds regardless the double signs appearing in the numerical coefficients Eq.(\ref{coeff}). For 
\beq
\Delta>0 \qquad \Rightarrow \qquad m_H<\frac{1}{3\sqrt{3}H(t)}
\eeq
 there are three real distinct roots. For 
\beq
\Delta=0 \qquad \Rightarrow \qquad m_H=\frac{1}{3\sqrt{3}H(t)}
\eeq
all the roots are still real with one of them being repeated. For 
\beq 
\Delta<0 \qquad \Rightarrow \qquad m_H>\frac{1}{3\sqrt{3}H(t)}
\eeq
there is one real and two complex conjugate roots \cite{stegun}. For computing the solutions of a cubic equation, it is convenient to introduce the following notation:
\begin{eqnarray}
&& {\bar B}:=\frac{B}{A}\,, \quad {\bar C}:=\frac{C}{A}\,, \quad {\bar D}:=\frac{D}{A}\\
&& Q:=\frac{{\bar C}}{3}-\frac{{\bar B}^2}{9}\,, \quad R:=\frac{{\bar B}{\bar C}-3{\bar D}}{6}-\frac{{\bar B}^3}{27}\,, \quad S_{1,2}:=[R\pm\sqrt{Q^3+R^2}]^{1/3}
\end{eqnarray}
in terms of which the Cardano's formula provides the three roots of Eq.(\ref{cgen}) as \cite{stegun}:
\beq
\label{sol3rd}
\chi_1=S_1+S_2-\frac{{\bar B}}{3}\,, \qquad \chi_{2,3}=-\frac{S_1+S_2}{2}-\frac{{\bar B}}{3}\pm \frac{i\sqrt{3}}{2}(S_1-S_2)\,,
\eeq
in which $i^2=-1$. Moreover \cite{stegun}:
\beq
\chi_1 \cdot \chi_2 \cdot \chi_3=-\frac{D}{A}=-\frac{m^3_{\rm H}}{8}<0\,.
\eeq
Therefore we can find three real negative roots (none of which with physical interpretation), or one negative and two positive roots. The latter will be interpreted as the black hole event horizon (the smaller root), and as the cosmological horizon (the larger root).

Ref. \cite{faraoni3} explored the possibilities of having one or multiple McVittie horizons for a mixture of dust and a cosmological constant, while in this manuscript we are interested in investigating these possibilities for a pure dark energy universe whose equation of state involves two free parameters (connected to the adiabatic speed of sound and the strength of interactions between the molecules constituting the fluid) as an extension of the case of the Schwarzschild-(anti-)de Sitter black hole studied in \cite{loc3}.

For setting up a meaningful system of equations we must write down the evolution equation for the scale factor $a(t)$ of the universe (or equivalently for its Hubble function) entering Eq.(\ref{3rd}).
The equations we need can be reduced to the Friedmann equation (mixed-rank time-time component of the field equations), the acceleration equation ($rr$, $\theta\theta$ and $\phi\phi$ components), and the equations accounting for the energy conservation (Bianchi identity):
\begin{eqnarray}
\label{eqtt}
&& H^2  \,=\,\left(  \frac{\dot a}{a}  \right)^2  \,=\,  \frac{\rho  }3 \\
\label{evH1}
&& 2\ddot \chi(2\chi+m_H)+\chi[2\chi(H^2+p)-m_H(5H^2 +p)]=0\\
\label{bianchie}
&& \dot \rho=-\frac{3H(2\chi-m_H)}{2\chi+m_H}(\rho+p) \\
\label{bianchip}
&& p'=\frac{4am_H(\rho+p)}{m_H^2 -4\chi^2}\,,
\end{eqnarray}
in which an appropriate relation $p=p(\rho)$ will be introduced in the next section. (\ref{evH1}) is equivalent to
\beq
2\dot H (2\chi+m_H)+(2\chi-m_H)(3H^2+p)=0\,,
\eeq
or to
\beq
\label{Hdot}
\dot H =-\frac{2\chi-m_H}{2(2\chi+m_H)}(\rho+p)\,,
\eeq
where we have used Eq.(\ref{eqtt}). We stress that Eq.(\ref{evH1}) correctly reduces to the Friedmann acceleration equation
\beq
\frac{\ddot a}{a}=-\frac{\rho+3p}{6}
\eeq
in the limit $m_H \to 0$. In this limit Eq. (\ref{bianchip}) provides as well $p(t,r)=p(t)$ as expected for a homogeneous and isotropic universe.

\section{Introducing the dark energy} \label{sIII}
Well-posed evolution equations for the McVittie manifold requires us to fix, a priori, a thermodynamic relation between the pressure and the energy density permeating the spacetime. We start by noticing that Eq.(\ref{eqtt}) implies that the energy density is spatially homogeneous, i.e. $\rho(t,r)=\rho(t)$. Then, to fulfill Eq.(\ref{bianchip}) we follow \cite{nandra1,nandra2} and write the pressure as
\beq
\label{pressurer}
p(t,r)=\rho(t)\left[(1+\omega(t))\frac{2\chi+m_H}{2\chi -m_H}F(t)-1\right]\,,
\eeq
with $F(t)$ and $w(t)$ being two arbitrary functions. Choosing $F(t)=1$ we can interpret $\omega (t)=p_\infty(t)/{\rho(t)}$ as the effective equation of state parameter at spatial infinity ${\chi}/{m_H}\gg 1$. Therefore Eq.(\ref{Hdot}) can be recast as
\beq
\label{HHH}
\dot H(t)=-\frac{\rho(1+\omega(t))}{2}=-\frac{\rho(t)+p_\infty(t)}{2}=-\frac{3H^2(t)+p_\infty(t)}{2}\,,
\eeq
from which the spatial homogeneity of the Hubble function appears more transparently. Similarly, Eq.(\ref{bianchie}) can be recast as
\beq
\dot \rho=-3H\rho(1+\omega(t)),
\eeq
showing that far away from the central massive object the energy conservation equation can be reduced to the same one which characterizes the Friedmann model  \cite{peebles}.

We connect the pressure to the energy density in the dynamical equations of the previous section assuming that the expansion of the  universe is driven by a dark energy fluid modeled according to the equations of state which are known under the names of Redlich-Kwong \cite{reos1}, (Modified) Berthelot \cite{reos2}, Dieterici \cite{reos3}, and Peng-Robinson \cite{reos4}. They read respectively as follow:
\begin{eqnarray}
\label{eos1}
p_\infty(t)&=& \frac{1- (\sqrt{2} -1) \alpha \rho}{ 1- (1-\sqrt{2}) \alpha \rho}\beta \rho \,, \\
\label{eos2}
p_\infty(t)&=& \frac{\beta \rho}{1+ \alpha \rho} \,, \\
\label{eos3}
p_\infty(t)&=& \frac{\beta \rho e^{2(1- \alpha \rho)}}{2- \alpha \rho} \,, \\
\label{eos4}
p_\infty(t)&=& \frac{\beta \rho}{1-\alpha \rho} \left[1-\frac{  (c_a/c_b)  \alpha \rho}{(1+ \alpha \rho)/(1- \alpha \rho) +  \alpha \rho}  \right]  \,, \qquad c_a\simeq 1.487\,, \,\, c_b \simeq 0.253 \,.
\end{eqnarray}

Ref. \cite{capo} discusses the applicability of such equations of state for the modeling of the dark energy in cosmology, and in particular its appendix reviews their microscopic foundations and their main thermodynamical properties focusing the attention on the possibilities of having a phase transition in these nonideal  fluids. We remark that the positive parameter $\alpha$ quantifies the deviations from a nonideal fluid behavior accounting for the internal interactions within the fluid molecules, because  all the above cases  reduce to a one-parameter ideal equation of state $p \sim \rho$ when $\alpha \to 0$. Moreover,  the parameter $\beta$  is connected to the adiabatic speed of sound $c^2_s={\partial p_\infty}/{\partial \rho}$ (as easily read off from the limit at small $\alpha$). The limiting cases of $\alpha=0$ and $\beta=\pm 1$ correspond to  a stiff fluid (which is equivalent to a massless scalar field \cite{cor1,cor2}),  and a cosmological constant, respectively.  These types of matter are relevant in the early and late-time cosmology respectively \cite{wei}. The case $\beta=0$ corresponds to dust (for example pressure-less dark matter).

Using the Friedmann equation, Eq.(\ref{eqtt}), we understand that the equations of state $p_\infty=p_\infty(\rho)$ can be rewritten in the more convenient form $p_\infty=p_\infty(H)$ as:
\begin{eqnarray}
\label{eos1a}
p_\infty(t)&=& 3H^2 \frac{1- 3(\sqrt{2} -1) \alpha H^2}{ 1- 3(1-\sqrt{2}) \alpha H^2}\beta  \,, \\
\label{eos2a}
p_\infty(t)&=& \frac{3\beta H^2}{1+3 \alpha H^2} \,,  \\
\label{eos3a}
p_\infty(t)&=& \frac{3\beta H^2 e^{2(1- 3\alpha H^2)}}{2- 3\alpha H^2} \,,  \\
\label{eos4a}
p_\infty(t)&=& \frac{3\beta H^2}{1-3\alpha H^2} \left[1-\frac{  3(c_a/c_b)  \alpha H^2}{(1+ 3\alpha H^2)/(1- 3\alpha H^2) +  3\alpha H^2}  \right]  \,,
\end{eqnarray}
for Eqs.(\ref{eos1})-(\ref{eos4}) respectively.

\subsection{Deceleration parameter}
The McVittie Universe (\ref{metric}) is shear-free and vorticity-free, with vanishing curvature parameter, but it admits nontrivial Hubble function, matter parameter, and acceleration vector as follows:
\begin{eqnarray}
H&=&  \sqrt{\frac{\rho}{3} } \\
\Omega_{\rm m}&=&  1 \\
\dot u^\alpha&=&\frac{64 r^4 m_H a^3(t)}{(2 a(t) r + m_H)^5 (2 a(t) r - m_H)} \delta_r^\alpha\,.
\end{eqnarray}
Thus, the deceleration parameter is\footnote{We refer the reader to eq. (59) in \cite{cosmo} for details on how the deceleration parameter must be computed for a non-Friedmann universe.}: 
\beq
q= \frac{1}{2}\left[1 +3\frac{p(t,r)}{ \rho(t)}\right].
\eeq
We stress that this result should not be interpreted in a Friedmann-like perspective because the McVittie universe is inhomogeneous and therefore the {\it effective equation of state parameter} $p(t,r)/ \rho(t)$ is indeed space-dependent. Therefore, the following three important aspects must be commented on. First of all, the direction of the energy flow from or to the cosmic background does not influence the sign of the deceleration parameter.  Then, a negative deceleration parameter necessarily requires an exotic fluid (i.e. with a negative pressure $p(t,r)$); this is a qualitatively different case than the inhomogeneous Stephani universe in which the acceleration vector can fully account for a realistic deceleration parameter without the need of any dark-energy-like fluid \cite{hashemi}. Finally, it appears that a large value of the Hawking-Hayward quasi-local mass can lead to a negative pressure (and therefore a negative deceleration parameter) via eq. (\ref{pressurer}) even in the case of positive $\beta$.

\subsection{Cosmographic analysis in the far field limit}

\begin{figure}
\begin{center}
    $
    \begin{array}{cc}
{\includegraphics[scale=0.25, angle=0]{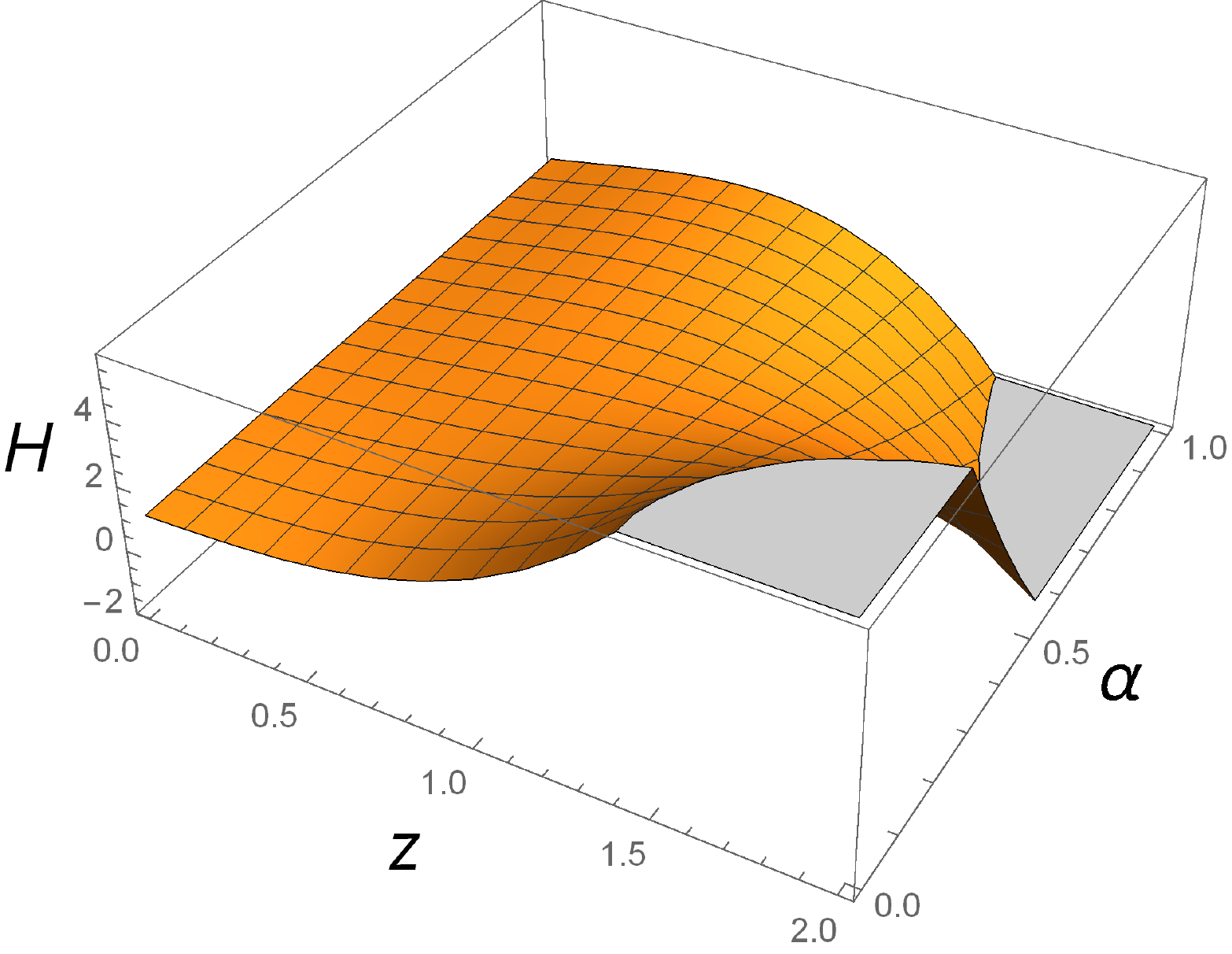}}
\hspace{3mm}
{\includegraphics[scale=0.25, angle=0]{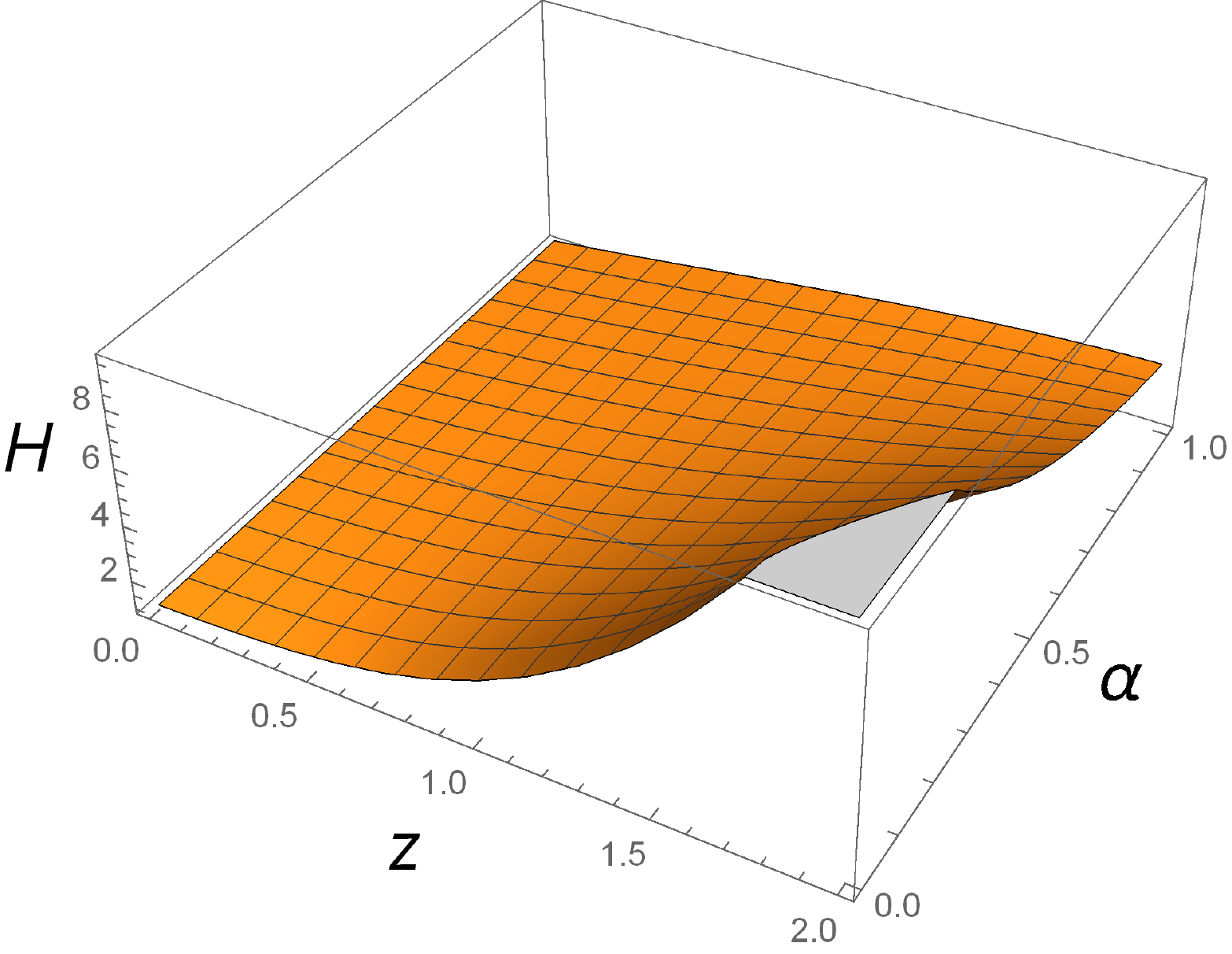}}
 \hspace{3mm} 
{\includegraphics[scale=0.25]{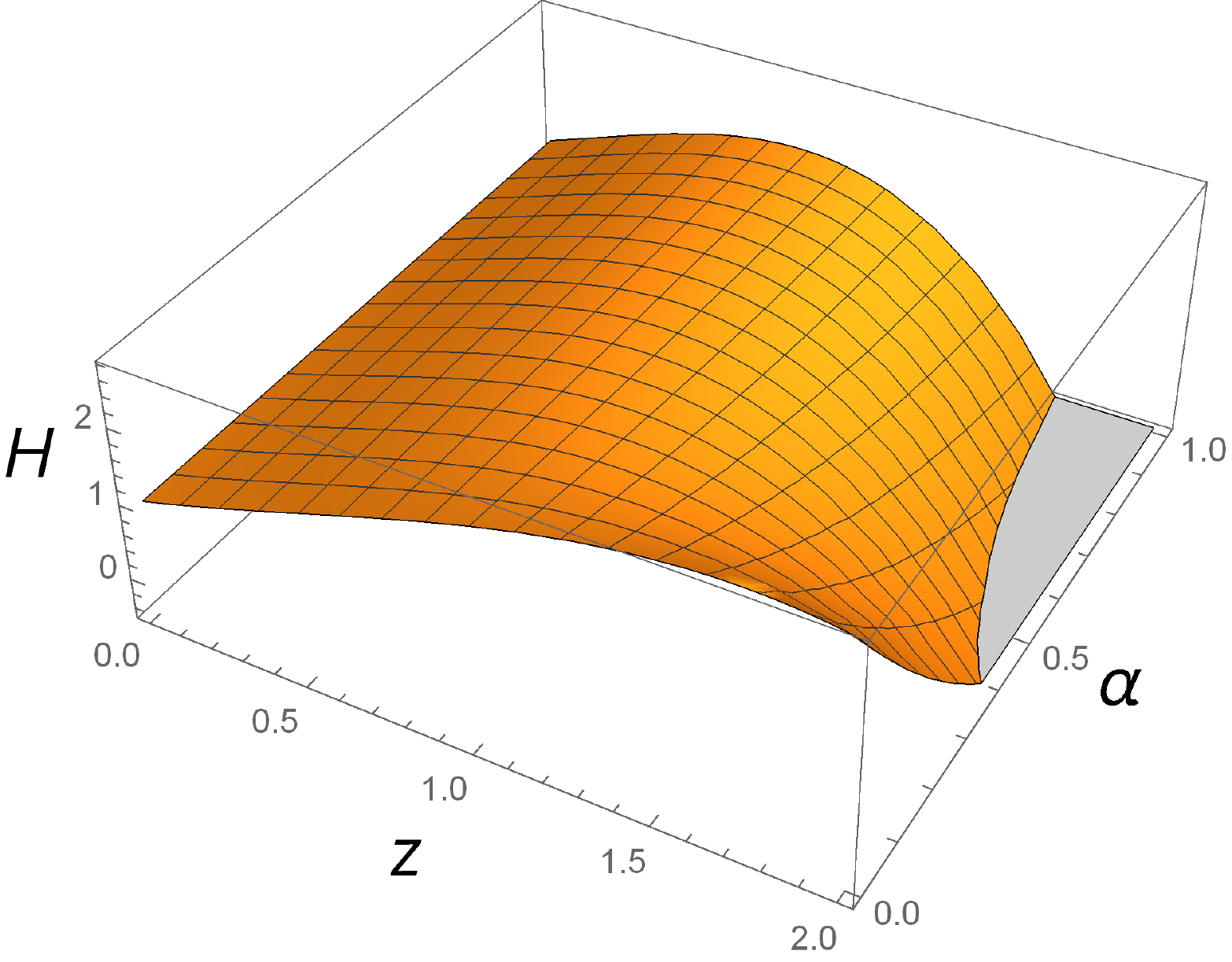}}
\hspace{3mm}
{\includegraphics[scale=0.25]{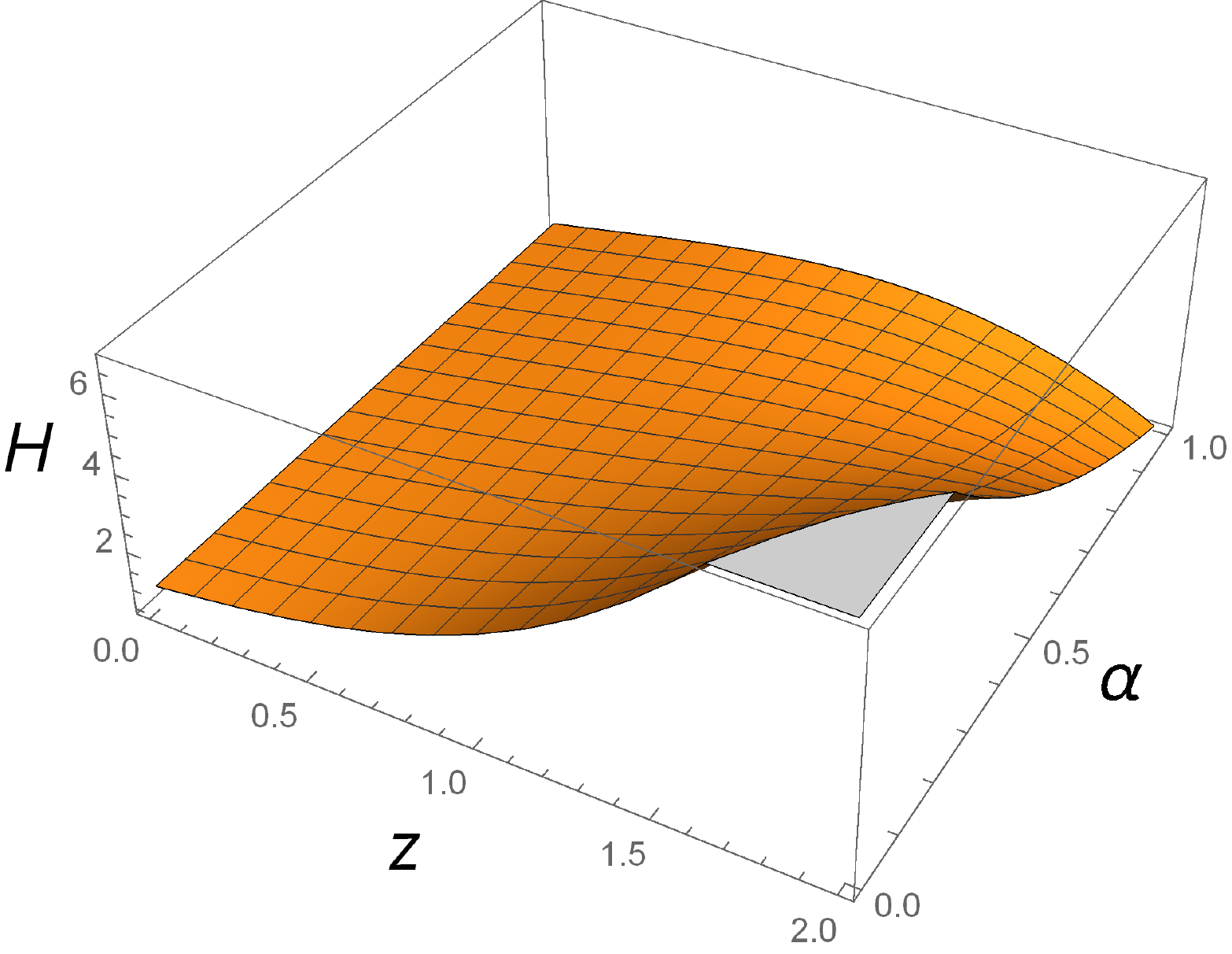}}\\
(a) \hspace{40mm} (b) \hspace{40mm} (c) \hspace{40mm} (d) \\[7mm]
{\includegraphics[scale=0.25, angle=0]{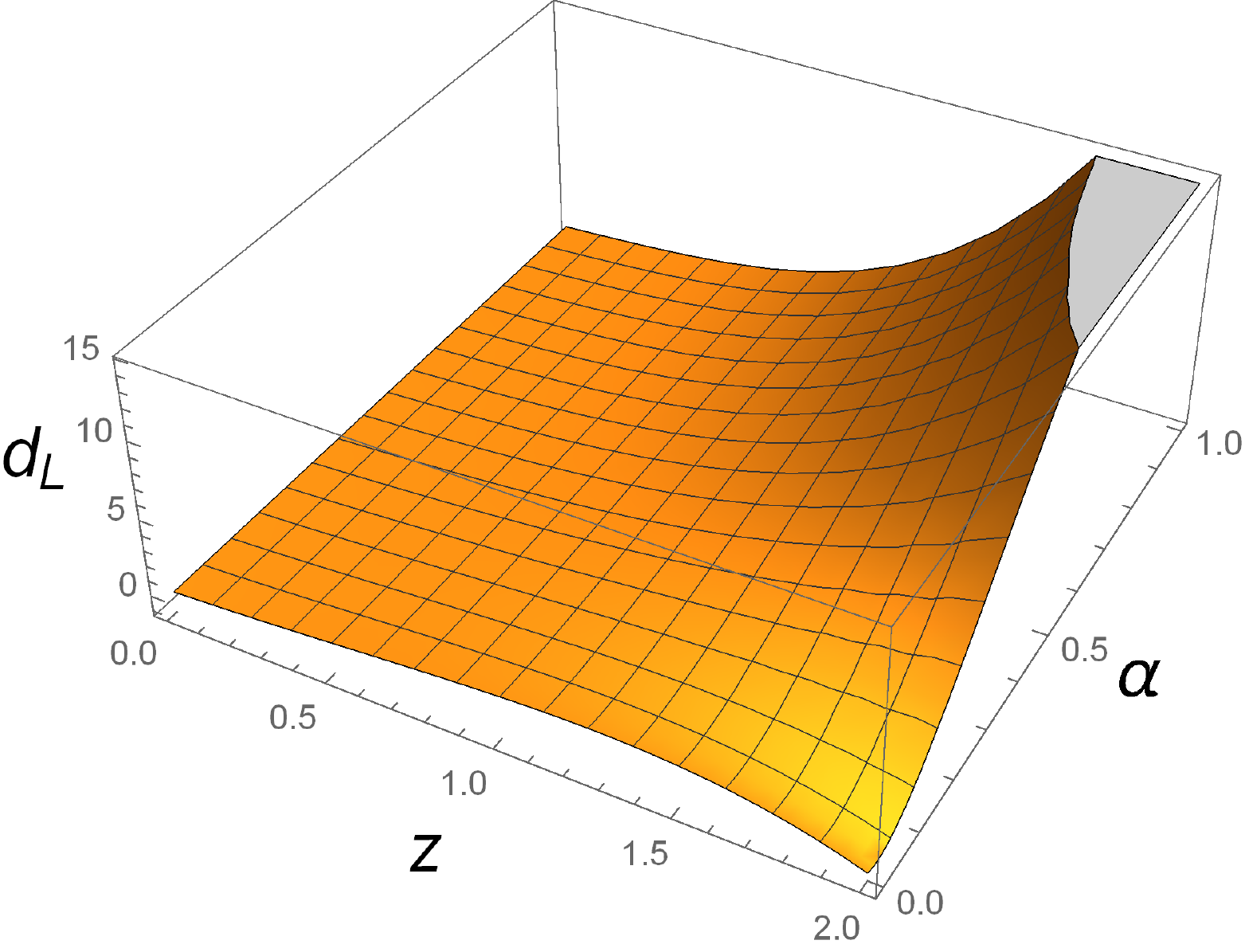}}
\hspace{3mm}
{\includegraphics[scale=0.25, angle=0]{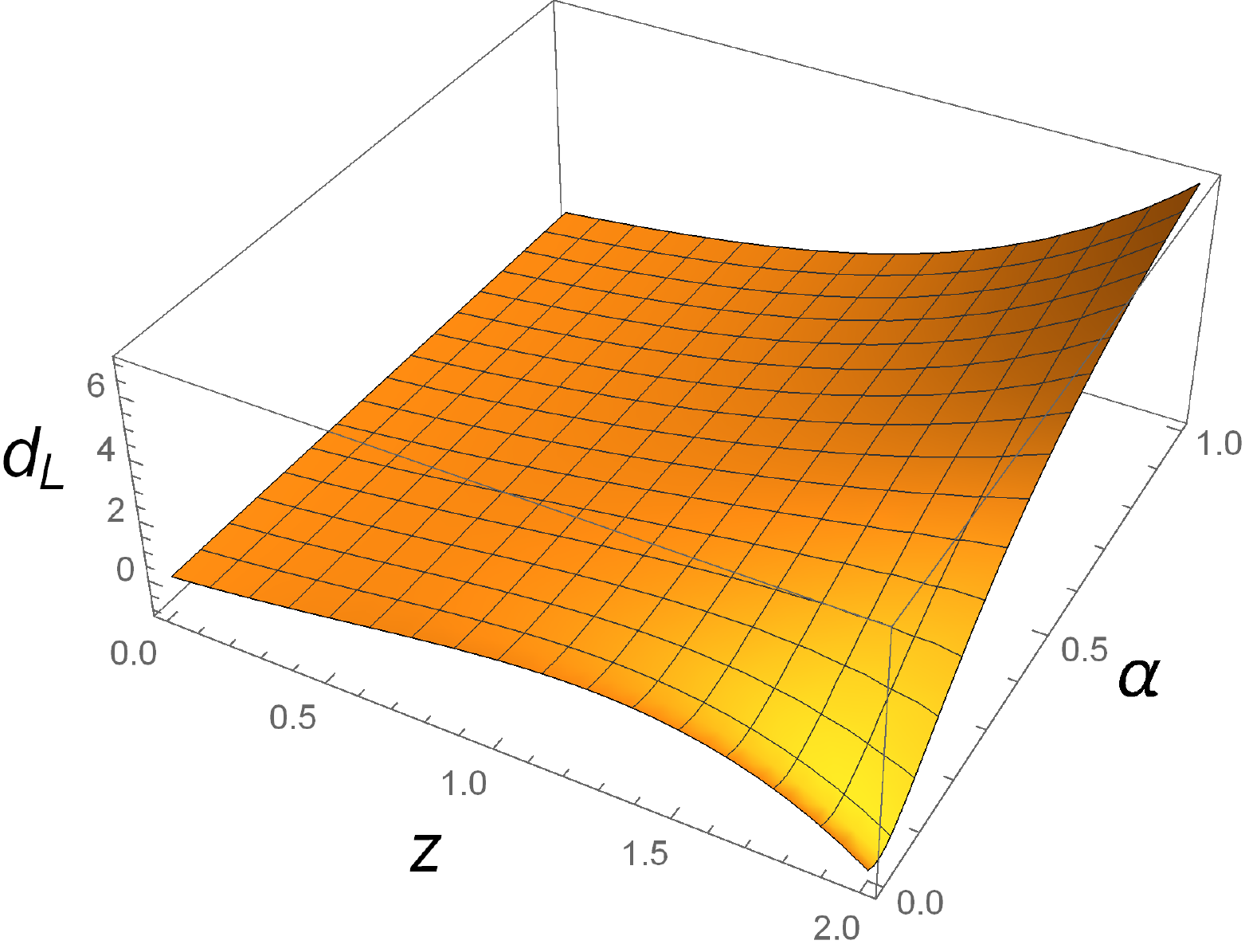}}
 \hspace{3mm} 
{\includegraphics[scale=0.25]{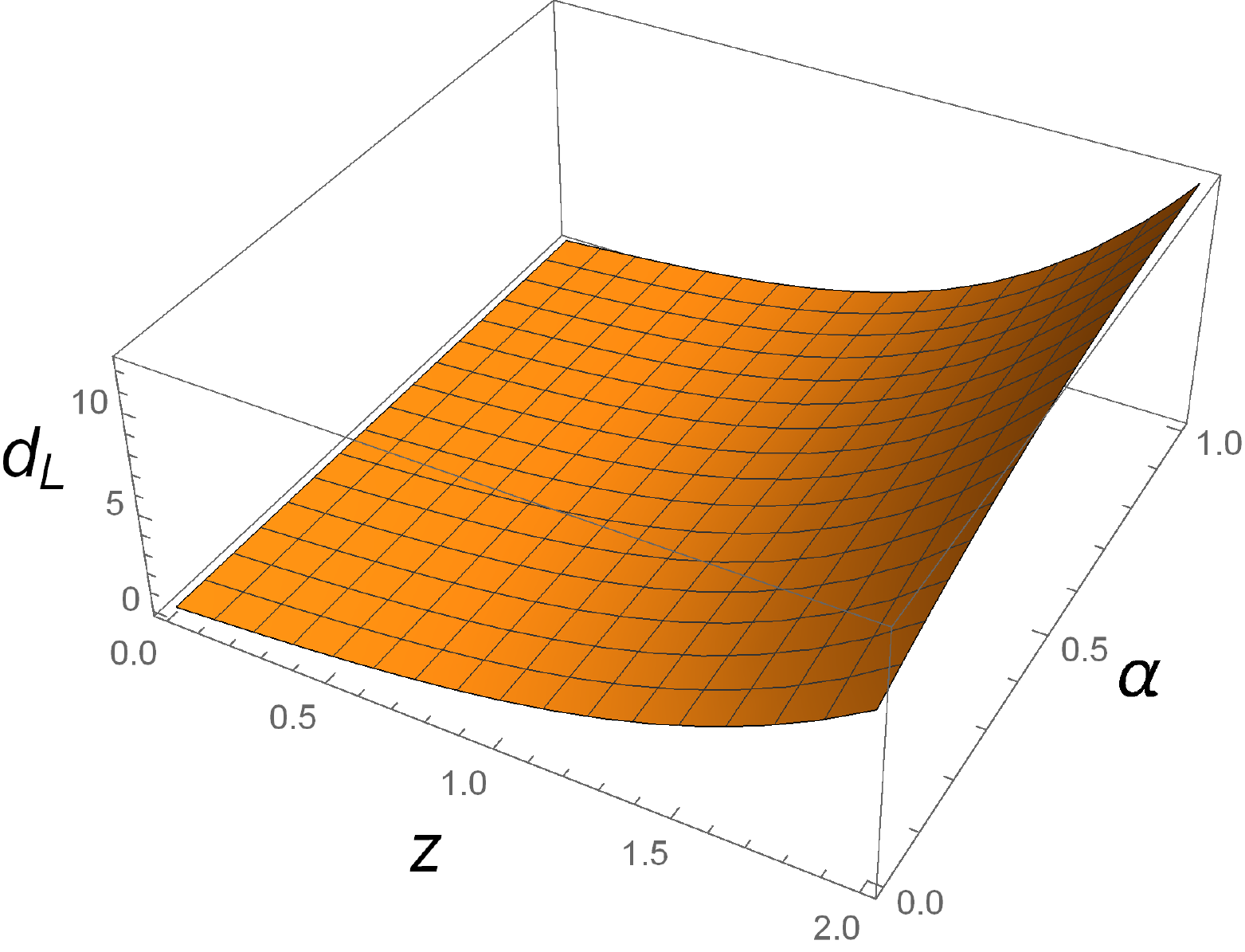}}
\hspace{3mm}
{\includegraphics[scale=0.25]{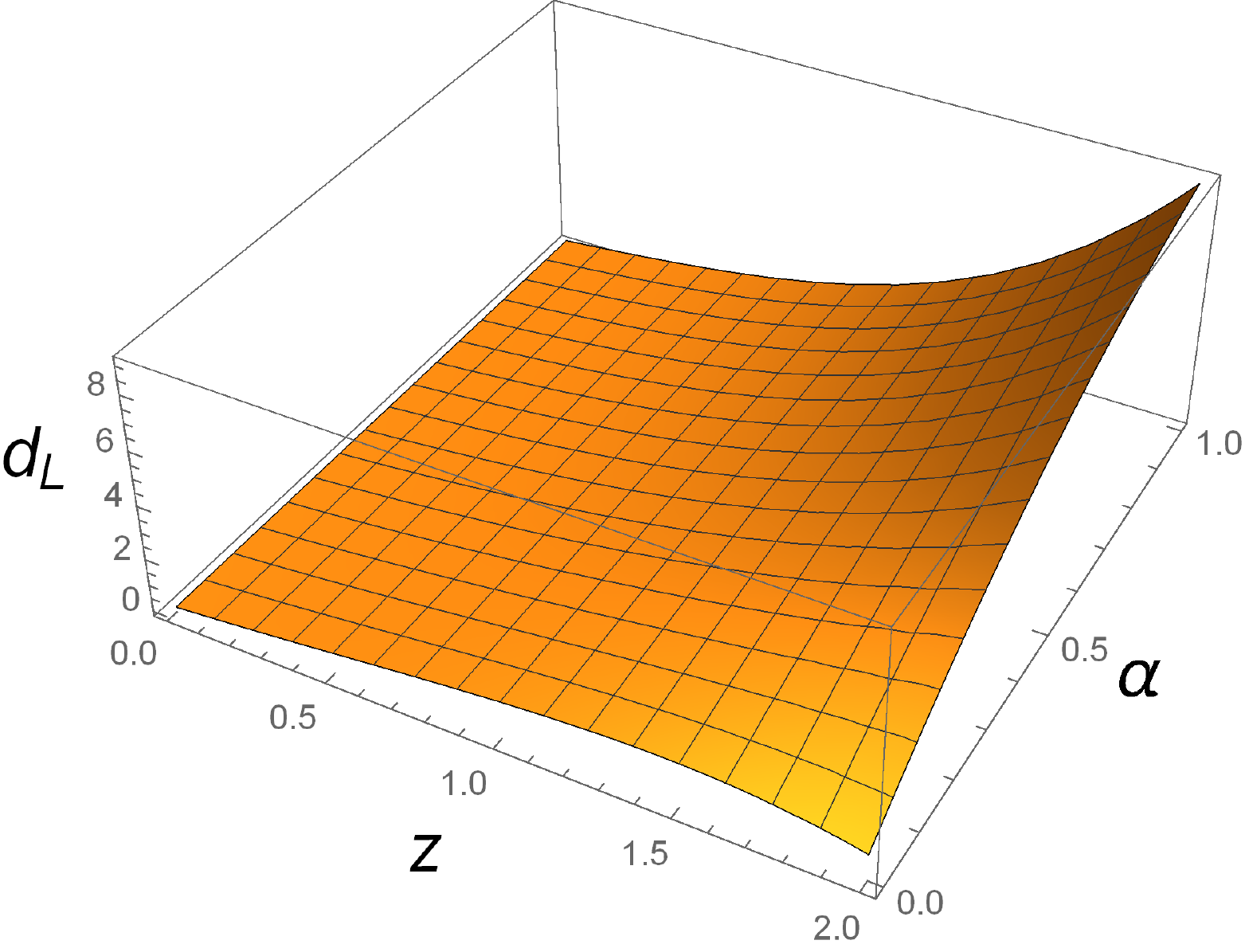}}\\
(e) \hspace{40mm} (f) \hspace{40mm} (g) \hspace{40mm} (h)
\end{array}
$
\end{center}
\caption{ The figure depicts the redshift evolution of the Hubble function and of the luminosity distance for the following fluid models:  Redlich-Kwong in panels (a) and (e), (Modified) Berthelot in panels (b) and (f), Generalized Chaplygin  in panels (c) and (g), and Anton-Schmidt in panels (d) and (h). We have adopted units such that $H_0=1$ (which implies $\rho_0=3$ through the Friedmann equation), and we have fixed $\beta=-0.75$, and $\rho_*=1$.
}
\label{comp}
\end{figure}

In the far field limit the Einstein equations of the McVittie spacetime are reduced to the usual Friedmann equations
\beq
\label{frcosmo}
H^2=\rho/3\,, \qquad \dot \rho=-3H(\rho +p)\,.
\eeq
In this regime we can adopt the cosmographic expansions for the luminosity distance and for the Hubble function in terms of the redshift $z$ \cite{visser1,visser2,visser3,visser4,visser5}:
\begin{eqnarray}
\label{luminosity}
d_L(z)&\simeq &\frac{z}{H_0} \left[1+\frac{(1-q_0)z}{2}+\frac{(-1+q_0+3 q_0^2 +j_0)z^2}{6} +\frac{(2-2q_0-15 q_0^2 -15 q_0^3 +5 j_0 +10 q_0 j_0 +s_0)z^3}{24}  \right]\\
\label{hcosmo}
H(z)&\simeq& H_0 \left[1+(1+q_0)z+\frac{(j_0 - q_0^2)z^2}{2} +\frac{(3q^2_0 + 3q^3_0
- j_0(3 + 4q_0) - s_0)z^3}{6}\right]\,,
\end{eqnarray}
in which we have introduced the deceleration, jerk, and snap parameters  which can be evaluated using the Friedmann Eqs. (\ref{frcosmo}):
\begin{eqnarray}
q&:=&-1-\frac{\dot H}{H^2} \\
j&:=&\frac{\ddot H}{H^3} -3q -2 \\
s&:=& \frac{ \dddot H}{H^4}+4j+3q(q+4)+6\,,
\end{eqnarray}
and a subscript $0$ indicates that the quantity is evaluated at the present time. The luminosity distance (\ref{luminosity}) can be measured through observations of type Ia supernovae \cite{riess1,riess2}, while the evolution of the Hubble function (\ref{hcosmo}) can be reconstructed using the distance to passively evolving galaxies \cite{riess3}. A direct computation delivers:
\begin{eqnarray}
q&=&\frac{1+3\omega_\infty}{2}  \\
j&=& 1+\frac{9 c_s^2 (1+\omega_\infty)}{2}  \\
s&=&  1+ \frac{27 (1+\omega_\infty)^2}{2}\left(1-\frac{ c_s^2 }{2}-\frac{\partial^2 p_\infty}{\partial \rho^2}  \right) + 9(1+\omega_\infty)\left( 2c_s^2 -\frac{3(1+c_s^2)^2}{2}  \right) \,,
\end{eqnarray}
where 
\beq
c_s=\sqrt{\frac{\partial p_\infty}{\partial \rho} }
\eeq
 is the adiabatic speed of sound of the cosmic fluid at spatial infinity. Refs. \cite{chap1,chap2,chap3} investigated the possibility of realizing {\it cosmic acceleration} adopting a single fluid two-parameter model in the framework of the Generalized Chaplygin Gas and of the Anton-Schmidt fluid. These two latter equations of state read as
 \begin{eqnarray}
 p&=& \frac{\beta}{\rho^\alpha}\\
  p&=& \beta \left(\frac{\rho_*}{\rho}\right)^\alpha \ln \frac{\rho}{\rho_*}\,,
\end{eqnarray}
respectively. Fig. (\ref{comp}) compares and contrasts the redshift evolution of the Hubble function and of the luminosity distance for the following fluid models: Redlich-Kwong in panels (a) and (e), (Modified) Berthelot in panels (b) and (f), Generalized Chaplygin in panels (c) and (g), and Anton-Schmidt in panels (d) and (h). We have adopted units such that $H_0=1$ (which implies $\rho_0=3$ through the Friedmann equation), and we have fixed $\beta=-0.75$, and $\rho_*=1$. The non-trivial behavior as function of the free parameter $\alpha$ can be used for discriminating between these cosmological models.

\section{Numerical analysis} \label{sIV}

\begin{figure}
\begin{center}
    $
    \begin{array}{cc}
{\includegraphics[scale=0.3, angle=0]{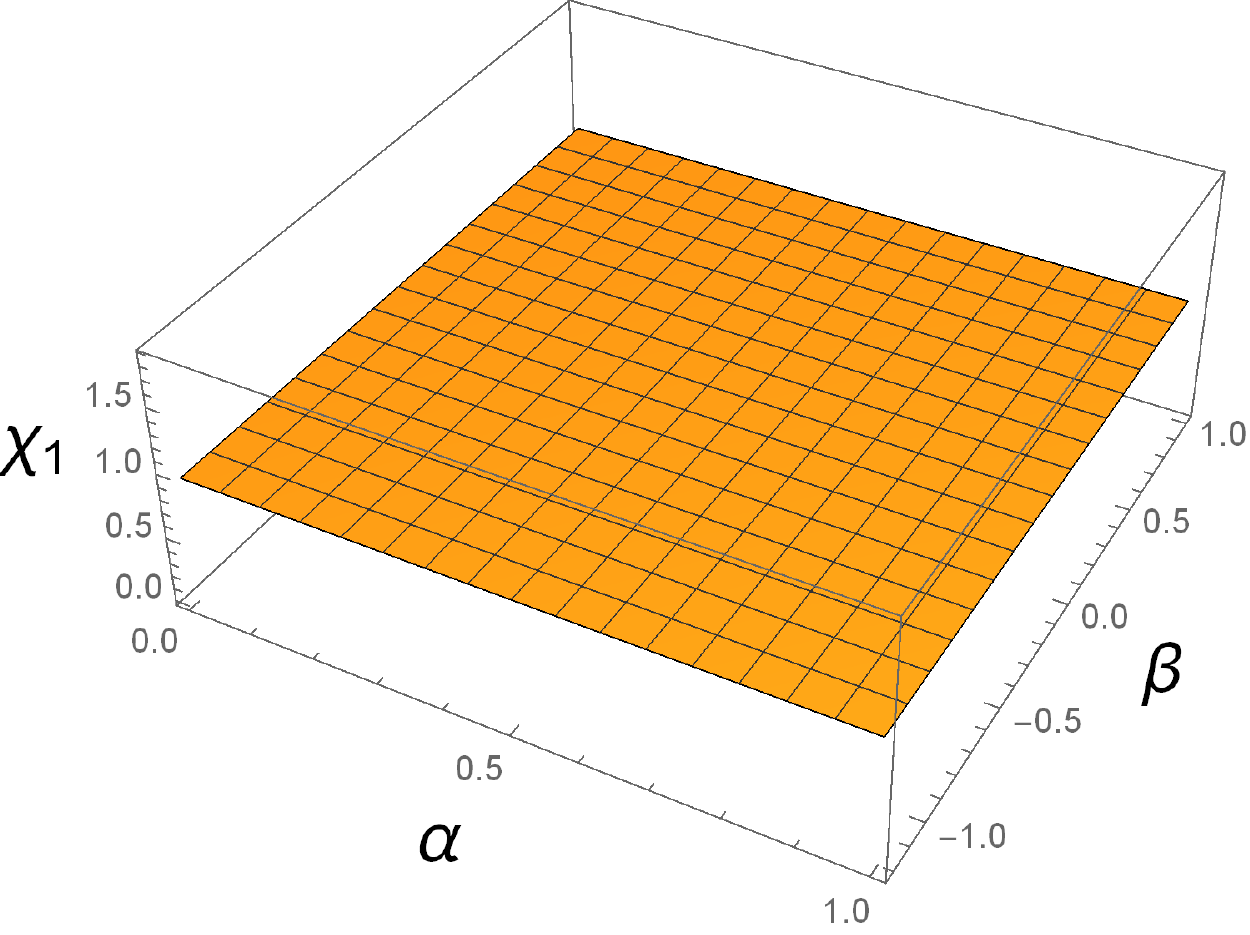}}
\hspace{3mm}
{\includegraphics[scale=0.3, angle=0]{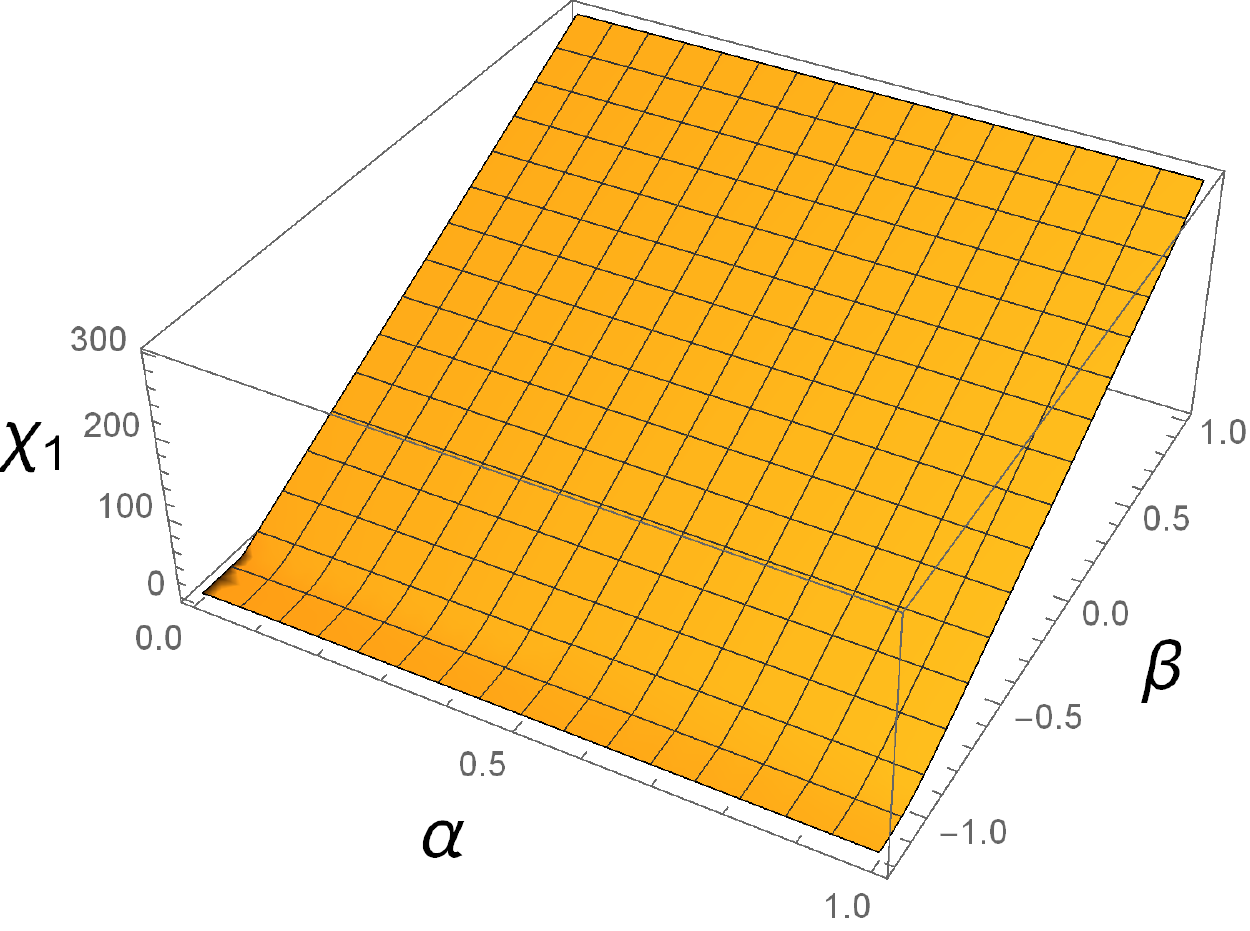}}
 \hspace{3mm} 
{\includegraphics[scale=0.3]{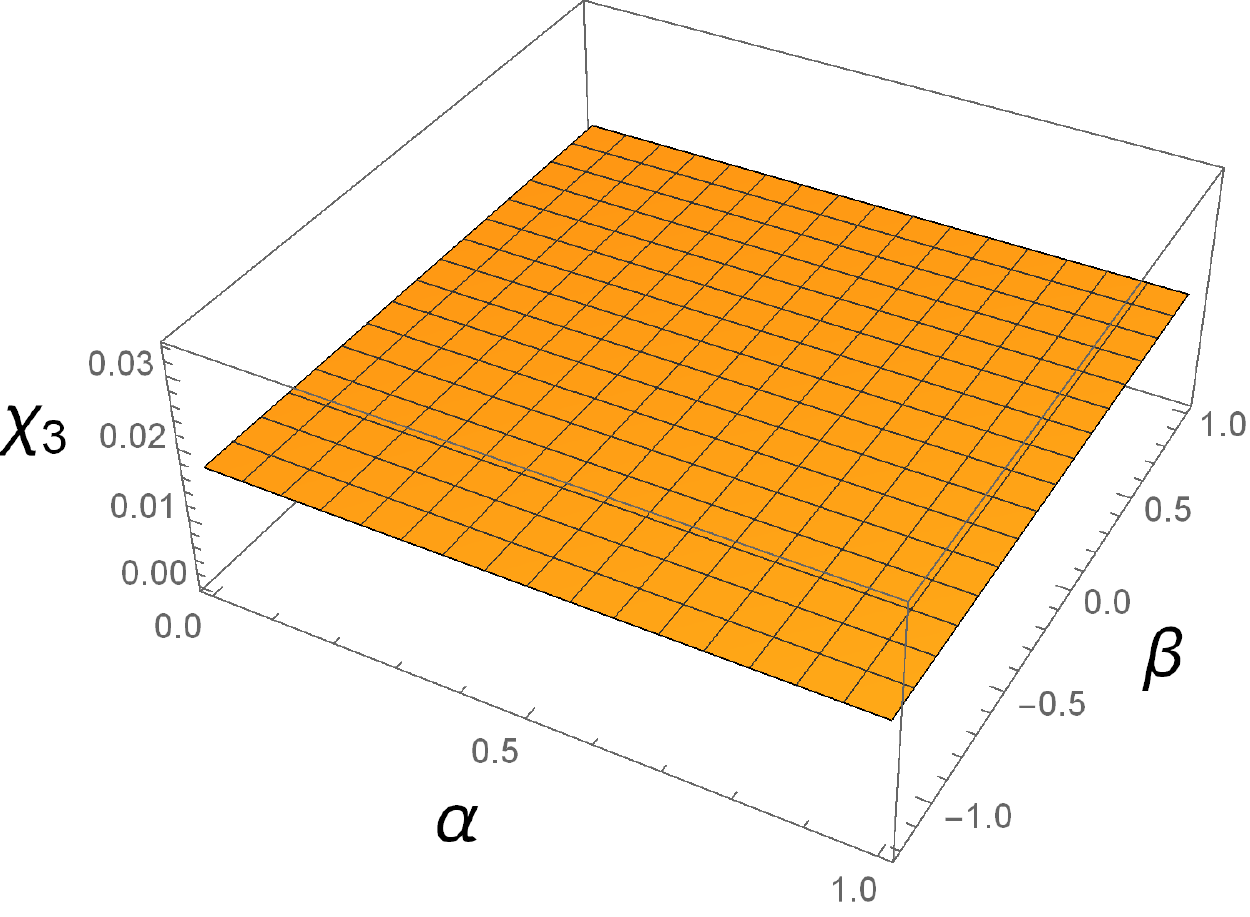}}
\hspace{3mm}
{\includegraphics[scale=0.3]{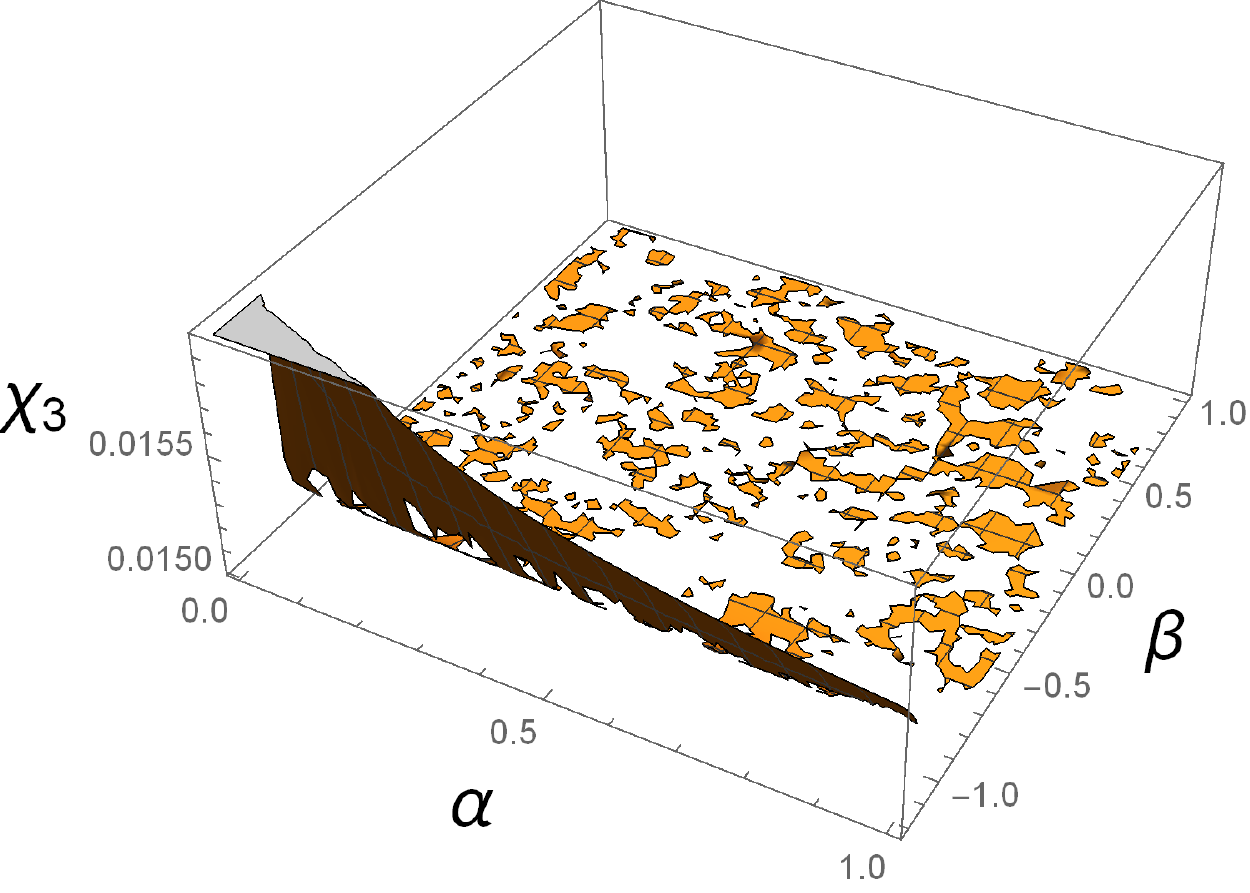}}\\
(a) \hspace{40mm} (b) \hspace{40mm} (c) \hspace{40mm} (d) \\[7mm]
{\includegraphics[scale=0.3, angle=0]{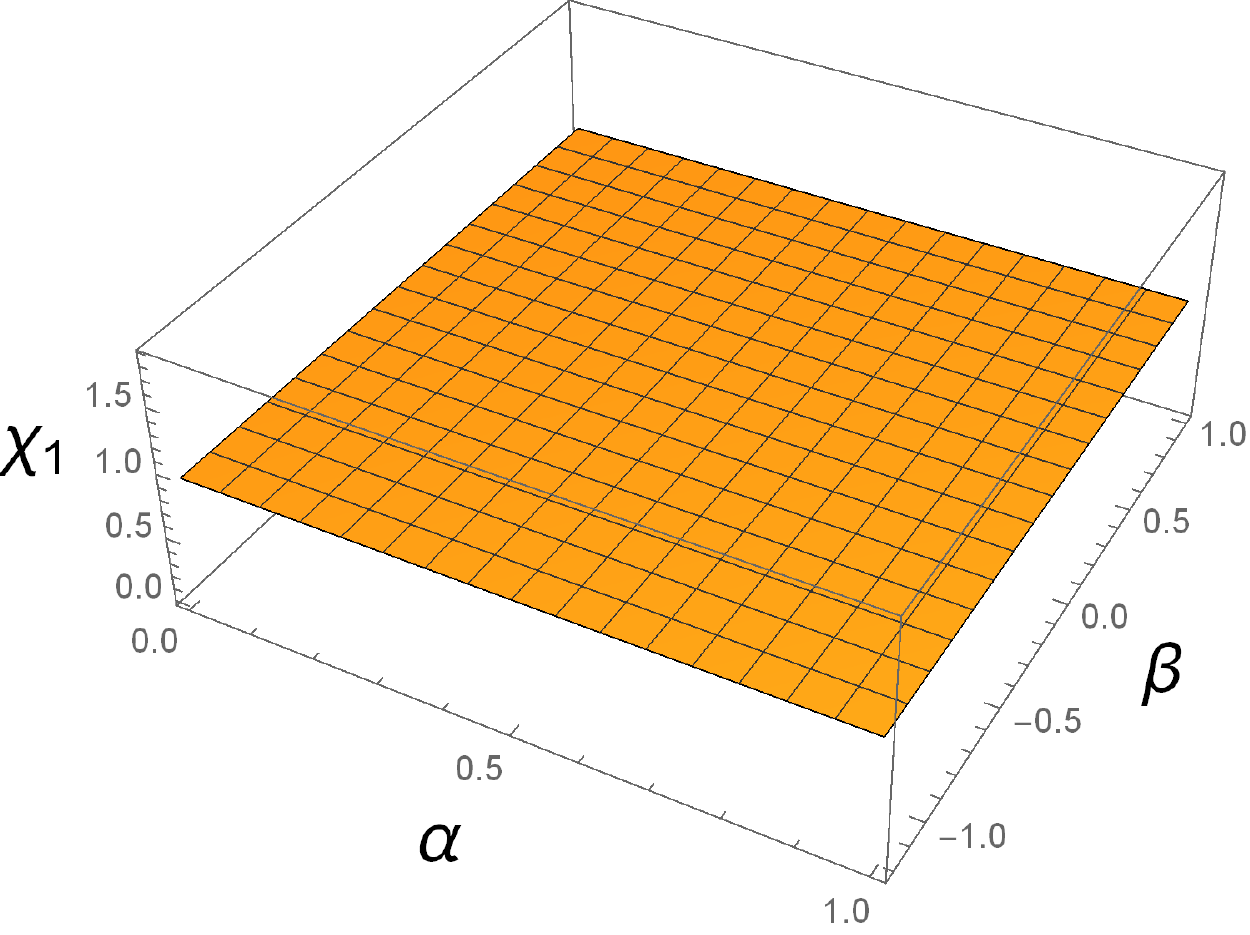}}
\hspace{3mm}
{\includegraphics[scale=0.3, angle=0]{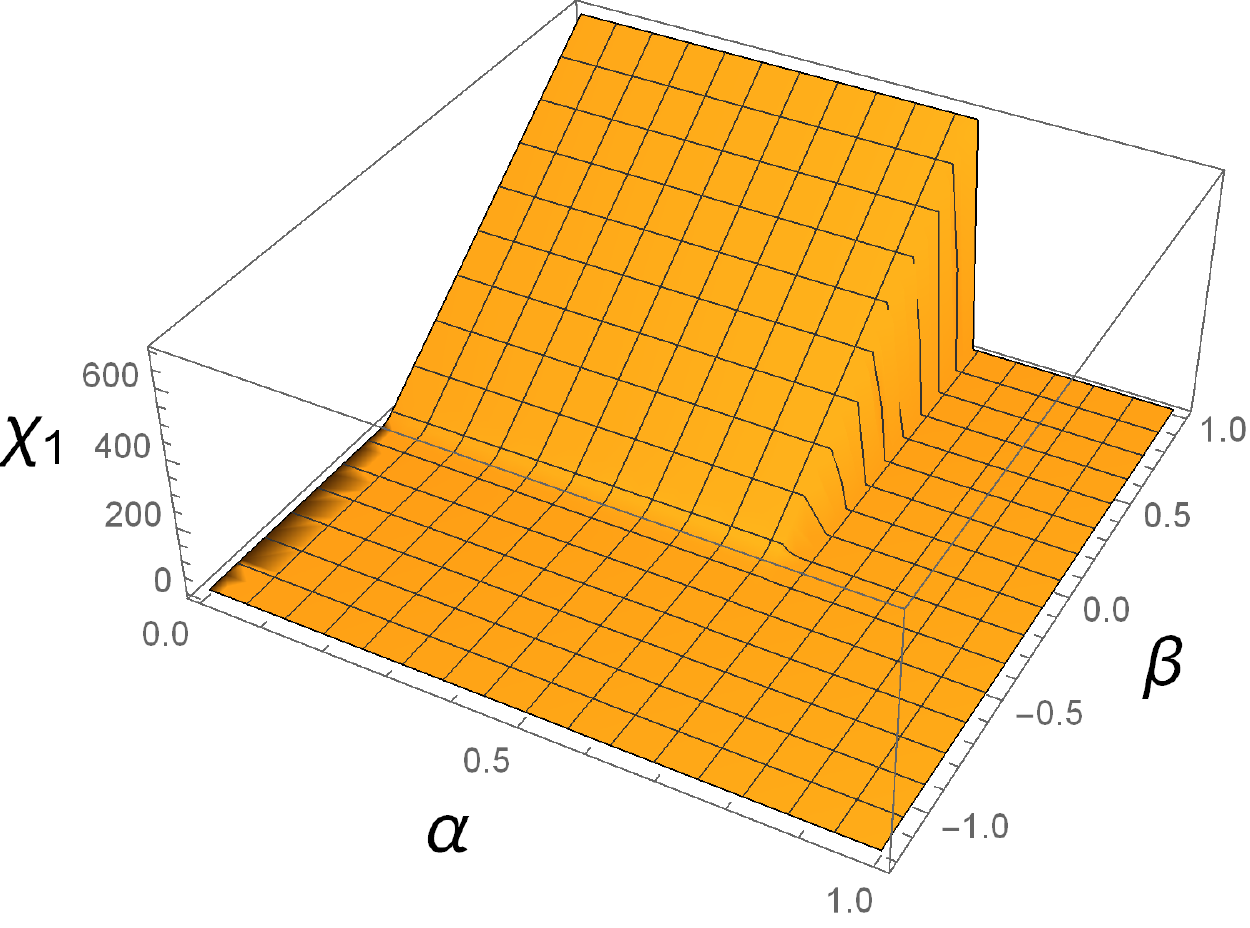}}
 \hspace{3mm} 
{\includegraphics[scale=0.3]{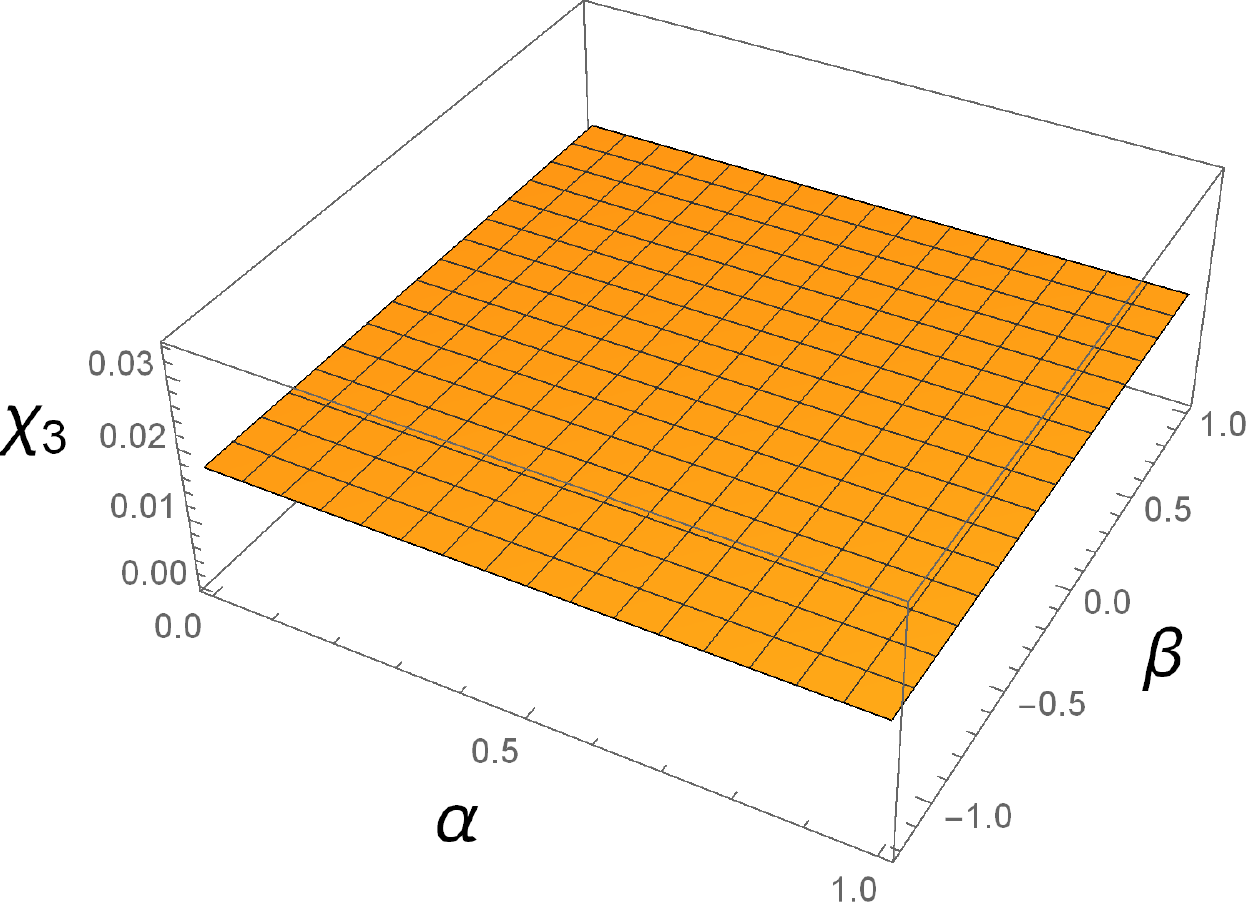}}
\hspace{3mm}
{\includegraphics[scale=0.3]{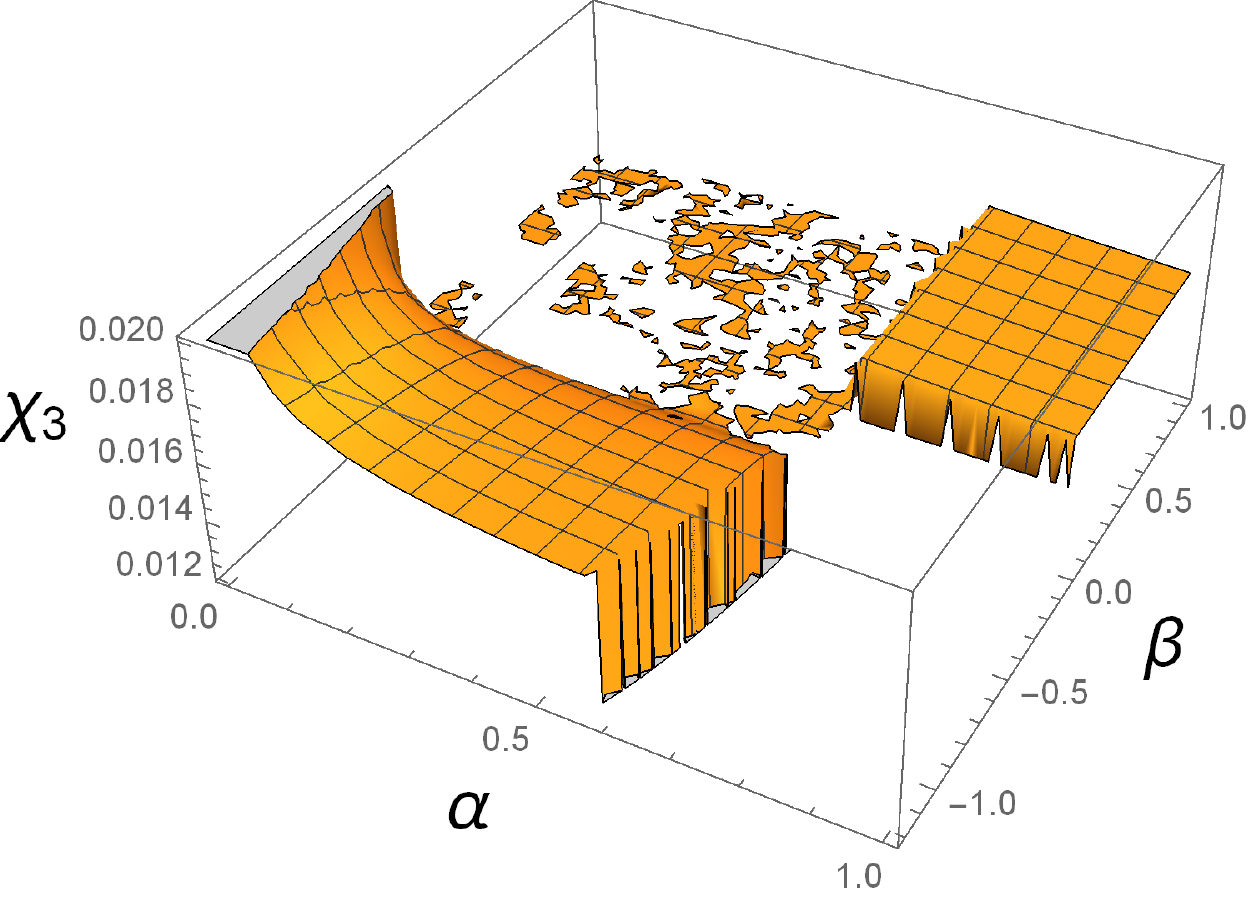}}\\
(e) \hspace{40mm} (f) \hspace{40mm} (g) \hspace{40mm} (h)
\end{array}
$
\end{center}
\caption{ The figure depicts the snapshots at time $t=1.0$ and $t=100.0$  for the evolution of the cosmological horizon $\chi_1$, and  of the event horizon $\chi_3$  in panels (a)-(b)-(e)-(f) and (c)-(d)-(g)-(h) respectively of the McVittie spacetime. The evolution of this universe is driven by the (Modified) Berthelot  fluid in the former case and by the Dieterici  fluid in the latter.
}
\label{fig1}
\end{figure}

\begin{figure}
\begin{center}
    $
    \begin{array}{cc}
{\includegraphics[scale=0.3, angle=0]{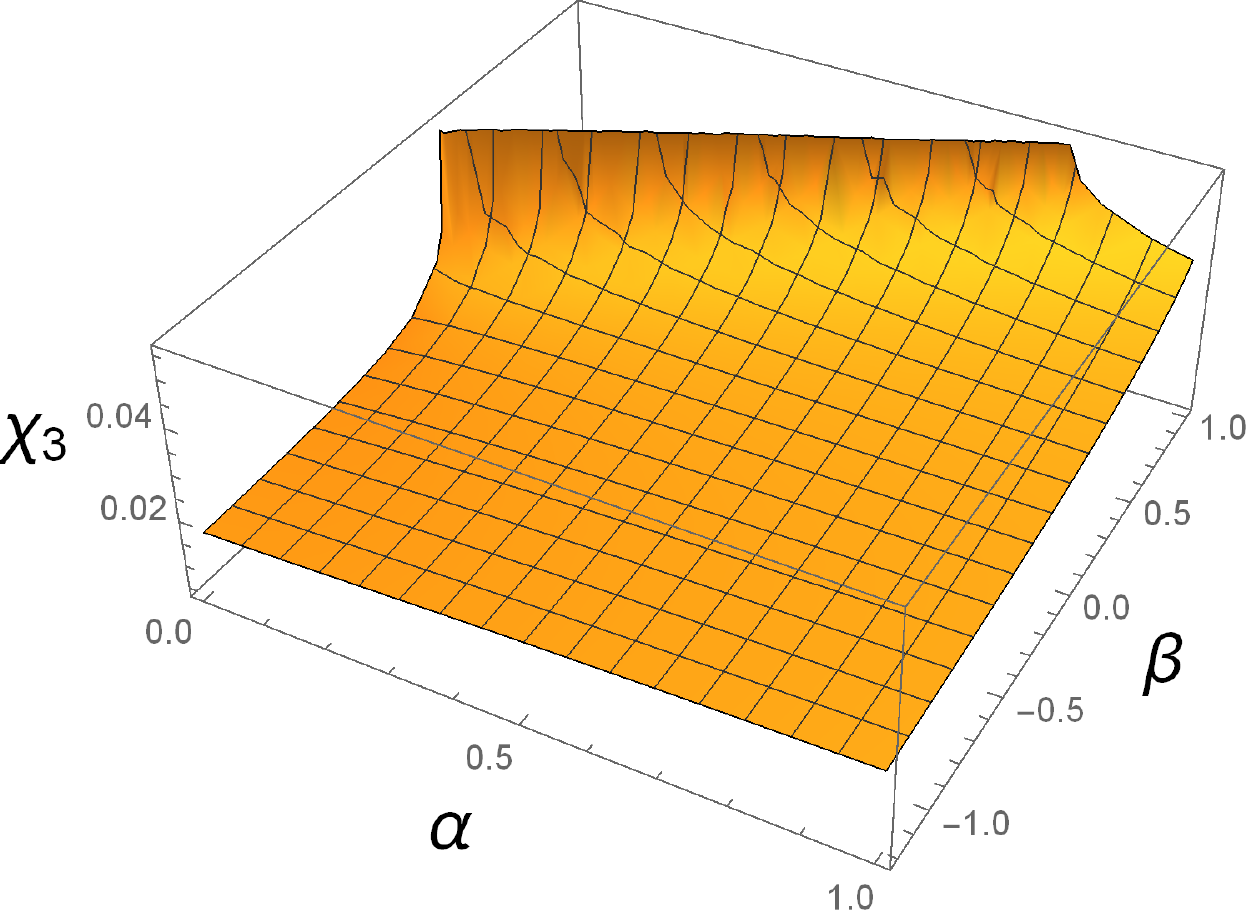}}
 \hspace{3mm} 
{\includegraphics[scale=0.3]{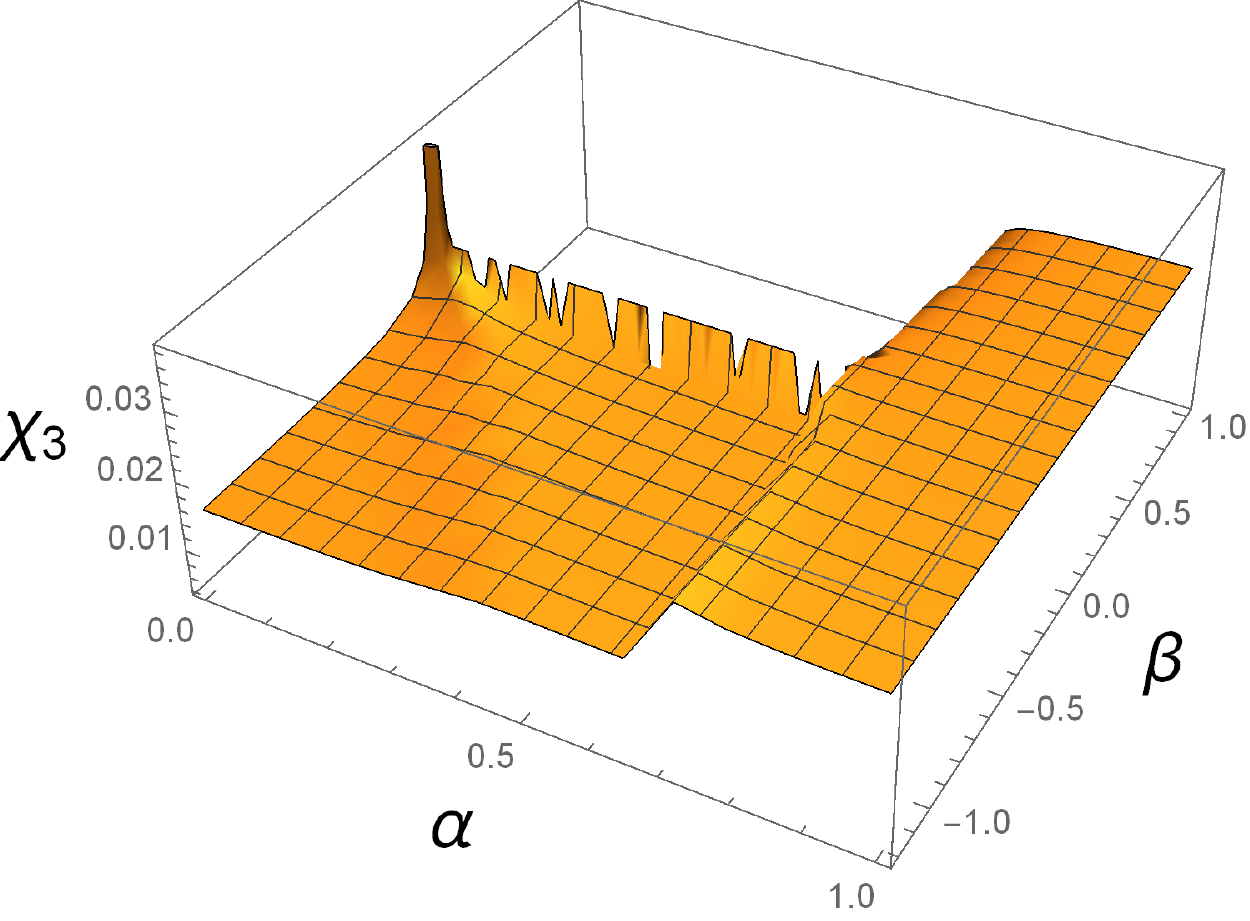}}
\\
(a) \hspace{40mm} (b) 
\end{array}
$
\end{center}
\caption{The figure depicts the snapshots at time $t=0.5$ of the McVittie event horizon for the  (Modified) Berthelot and Dieterici equations of state respectively  in panels (a) and (b).
}
\label{fig5}
\end{figure}

The locations for the McVittie horizons are constructed from Eq.(\ref{sol3rd}) with the following substitutions:
\begin{eqnarray}
&& {\bar B}=     \frac{3 m_H}{2}-\frac{1}{H}    \,, \quad S_{1,2}= \frac{2^{\frac{1}{3}}}{6H} \Big[4+m_H H[27(m_H  H-1) \pm 3\sqrt{3}\sqrt{27H^2 m^2_H-1} ]\Big]^{1/3},      \\
&& {\bar B}=     \frac{3 m_H}{2}+\frac{1}{H}    \,, \quad S_{1,2}= -\frac{2^{\frac{1}{3}}}{6H} \Big[4 + m_H H[27(m_H  H+1) \mp 3\sqrt{3}\sqrt{27H^2 m^2_H-1} ]\Big]^{1/3},
\end{eqnarray}
for the cases $\chi>m_H/2$ and $\chi<m_H/2$, respectively. In these solutions the time evolution of the Hubble function $H=H(t)$ is computed by integrating (\ref{HHH}), and then replacing the pressure at spatial infinity with formulas (\ref{eos1a})-(\ref{eos4a}).         
 Fig. (\ref{fig1}) displays the snapshots when $t=1.0$ and $t=100.0$ of the time evolution of the cosmological horizon $\chi_1$ in panels (a)-(b)-(e)-(f) and of the event horizon $\chi_3$  in panels (c)-(d)-(g)-(h) ($\chi_2$ being  negative, it does not carry any physical interpretation \cite{faraoni3}) for the  (Modified) Berthelot and for the Dieterici  modeling for the dark energy.  The dynamics of both the cosmological and black hole horizons for the Redlich-Kwong and  Peng-Robinson fluids are qualitatively the same,  and therefore are not explicitly shown.   Since our purpose is to investigate the effects that different types of dark energy or regular fluids have on the evolution of the McVittie horizons, we fix  throughout our analysis  as reference values  the Hawking-Hayward mass\footnote{In this paper we have  adopted units such that $c=1=8\pi G$.  The multiplication factor for converting units to SI units can be found in Appendix F of \cite{wald}, and they are the following: for mass one multiplies $8 \pi G c^{-2}\simeq 0.186 \cdot 10^{-25}$ m/kg, and for time one multiplies $c\simeq 3 \cdot 10^{8}$  m/s. For the Hubble function, its dimension is the inverse of  time, and for the quantity $\chi$, its dimension is that of mass, as can be seen from Eq.(\ref{3rd}).}  $m_H=0.03$  and the initial condition $H(t_0=1.0)=1.0$ for the numerical integration of (\ref{HHH}) for graphical convenience.  Fig. (\ref{fig5}) displays the snapshots at time $t=0.5$ (integrating the evolution of the system backwards) of the cosmological horizons $\chi_3$ for the  (Modified) Berthelot, and for the Dieterici  equations of state, respectively in panels (a) and (b).

\subsection{Discussion}

A number of regularity properties appear from the numerical analysis presented in Figs (\ref{fig2})-(\ref{fig5}).  Along the time evolution we considered, we have $\chi_1 > \chi_3$  for all the four configurations. Therefore, the former is interpreted as the cosmological horizon, while the latter is the black hole event horizon. The cosmological horizon exists for all the values of the parameters $\alpha \in [0,1]$, and $\beta \in [-1,1]$ all along the time evolution, while the event horizon can exist only for certain appropriate pairs ($\alpha$, $\beta$) that are compatible with having a real solution to the cubic equation locating the McVittie horizons via Eq.(\ref{3rd}). This behavior is not surprising because as a general property of a third degree algebraic equation, we discussed that one real solution is always guaranteed. A second real non-complex solution may or may not arise according to a time dependent relationship between the Hawking-Hayward quasi-local mass   and the Hubble function, implying that it can disappear in time. Moreover, the cosmological horizon $\chi_1$ is expanding in time, while the event horizon $\chi_3$ is shrinking. This behavior is compatible with the general property that the product and the sum of the three roots of a cubic equation must remain fixed \cite{stegun}. The cosmological horizon is attaining larger sizes for regular types of matter characterized by $\beta>0$, rather than for exotic matter with $\beta<0$ regardless the strength of non-linearities within the fluid quantified by the parameter $\alpha$. On the other hand, this is not the case for the event horizon which exhibits larger sizes for exotic matter and for ideal fluids with $\alpha \to 0$. We remark that all these properties hold regardless the modeling for the fluid we adopted in terms of the Redlich-Kwong, (Modified) Berthelot, Dieterici, or Peng-Robinson equations of state. However,  the latter two exhibits a singularity for the pressure for finite values of the energy density at $\rho=2/\alpha$ and at $\rho=1/\alpha$ respectively. This divergence influences the evolution of the Hubble function via (\ref{HHH}) and brings to a steepening of the evolution of both the cosmological and event horizons of the McVittie spacetime.

\section{Concluding remarks} \label{sV}

The physics accounting for the existence of black holes and the one explaining the global evolution of the Universe may seem unrelated at first sight because the former involves physics at relatively small scale, while the latter focuses on large scale effects. Just to mention one example, some current research on black holes is trying to picture the formation of an accretion disk on scales of 0.20 pc as in the galaxy NGC 3783 \cite{ngc}, while  cosmological literature is trying to clarify the role of spatial inhomogeneities on sizes of 150 Mpc \cite{millenium}. The spatial scales of these two phenomena differ from each other by 9 order of magnitude. For addressing the open issue of providing a direct astrophysical evidence of the existence of a black hole horizon which does not rely on the detection of the energy emitted by the particles falling into it \cite{BIN}, the Event Horizon Telescope provided the first physical image of an object of the size of the event horizon of a black hole located at the center of the Milky way \cite{telescope} (technically speaking the black hole shadow is larger than the horizon size, but it is of the same order of magnitude). Therefore, the analysis presented in this manuscript intends to complement the ongoing research in which interactions between a black hole and an evolving background cosmic fluid are accounted for \cite{babi1,babi2,babi3,babi4}.  The key improvement is that in our model the Einstein equations automatically accounts for the evolution of the mass of the black hole (thanks to the adoption of the McVittie spacetime metric) without the need of imposing by hand any  ad-hoc dynamical equation for the mass which is absorbing or emitting energy from or towards the background region assumed to be a Friedmann universe.  The ratio of growth of galaxies and supermassive black holes have remained constant during the past 11 Gyr suggesting a close interplay between the evolution of the two \cite{astrorev1,astrorev2,astrorev3,astrorev4}. The interpretation is that phenomena connecting the cosmic evolution of host galaxies and of their central black hole still occur at the present age of the Universe and dominate over the merging of black holes or stellar collapse as a mechanism for the formation of the central supermassive black hole. In particular, tracing the radio and X-rays signals emitted by the material infalling into the black hole horizon in a sample of 35 Active Galactic Nuclei it was possible to observationally reconstruct the time evolution of the mass of the central supermassive black hole as a function of the temperature (energy) of the accreting matter \cite{agn1,agn2,agn3}. In light of these measurements, a set of magneto-hydrodynamic simulations have been carried out \cite{magneto1,magneto2}, that our paper intends to complement from an analytical perspective through the adoption of the McVittie metric.  Following this latter line of thinking, in this manuscript we have investigated the effects that the dark energy modeling beyond a cosmological constant and that regular fluids carry on the evolution of the location of the horizon of a black hole. In particular, we showed that a regular vs. an exotic or an ideal vs. a non-ideal matter content influences in different ways the evolution of the cosmological and of the event horizon in a McVittie spacetime. Therefore, the consequences of adopting different  modelings for the large-scale physical phenomena cannot be ignored in the study of local small-scale physics \cite{vs1}. Our study complements the ones which instead focus on the opposite way of thinking, i.e. in clarifying the roles that small-scale effects have on the global expansion of the universe \cite{vs2}. Last but not least, our analysis about the location of the event  horizon can play a role in numerical relativity simulations about gravitational waves in which the excision method is adopted,  and information about the location of the event horizon (i.e. about the region to excise) are necessary \cite{loc1}.

\subsection*{Acknowledgement}
Y.C.O. thanks NNSFC (grant No.11705162) and the Natural Science Foundation of Jiangsu Province (No.BK20170479) for
funding support. D.G. acknowledges support from China Postdoctoral Science Foundation (grant No.2019M661944).

{}

\end{document}